\documentclass[aps,prl,superscriptaddress]{revtex4-2}
\usepackage{amsmath}	
\usepackage{graphicx}	
\usepackage{color}
\usepackage{gensymb}
\usepackage{braket}
\usepackage{hyperref}
\usepackage{upgreek}

\begin{document}
\Large

\title{\Large Nonlinear interactions of dipolar excitons and polaritons in MoS$_2$ bilayers}

\author{Charalambos Louca}
\email{clouca1@sheffield.ac.uk}
\affiliation{Department of Physics and Astronomy, The University of Sheffield, Sheffield S3 7RH, UK}

\author{Armando Genco}
\email{armando.genco@polimi.it}
\affiliation{{Dipartimento di Fisica, Politecnico di Milano, Piazza Leonardo da Vinci, 32, Milano, 20133, Italy}}

\author{Salvatore Chiavazzo}
\affiliation{Department of Physics, University of Exeter, Stocker Road, Exeter, EX4 4PY, UK}

\author{Thomas P. Lyons}
\affiliation{Department of Physics and Astronomy, The University of Sheffield, Sheffield S3 7RH, UK}
\affiliation{RIKEN Center for Emergent Matter Science, Wako, Saitama, 351-0198, Japan}

\author{Sam Randerson}
\affiliation{Department of Physics and Astronomy, The University of Sheffield, Sheffield S3 7RH, UK}

\author{Chiara Trovatello}
\affiliation{{Dipartimento di Fisica, Politecnico di Milano, Piazza Leonardo da Vinci, 32, Milano, 20133, Italy}}

\author{Peter Claronino}
\affiliation{Department of Physics and Astronomy, The University of Sheffield, Sheffield S3 7RH, UK}

\author{Rahul Jayaprakash}
\affiliation{Department of Physics and Astronomy, The University of Sheffield, Sheffield S3 7RH, UK}

\author{Kenji Watanabe}
\affiliation{Advanced Materials Laboratory, National Institute for Materials Science, 1-1 Namiki, Tsukuba, 305-0044, Japan}

\author{Takashi Taniguchi}
\affiliation{Advanced Materials Laboratory, National Institute for Materials Science, 1-1 Namiki, Tsukuba, 305-0044, Japan}

\author{Stefano Dal Conte}
\affiliation{{Dipartimento di Fisica, Politecnico di Milano, Piazza Leonardo da Vinci, 32, Milano, 20133, Italy}}

\author{David G. Lidzey}
\affiliation{Department of Physics and Astronomy, The University of Sheffield, Sheffield S3 7RH, UK}

\author{Giulio Cerullo}
\affiliation{{Dipartimento di Fisica, Politecnico di Milano, Piazza Leonardo da Vinci, 32, Milano, 20133, Italy}}

\author{Oleksandr Kyriienko}
\affiliation{Department of Physics, University of Exeter, Stocker Road, Exeter, EX4 4PY, UK}

\author{Alexander I. Tartakovskii}
\email{a.tartakovskii@sheffield.ac.uk}

\affiliation{Department of Physics and Astronomy, The University of Sheffield, Sheffield S3 7RH, UK}

%
%
%
%

\begin{abstract}

\end{abstract}

\maketitle

{\bf Nonlinear interactions between excitons strongly coupled to light are key for accessing quantum many-body phenomena in polariton systems\cite{Deng2010,Kasprzak2006,Christopoulos2007,Bhattacharya2014,AmoNat2009}. Atomically-thin two-dimensional semiconductors provide an attractive platform for strong light-matter coupling owing to many controllable excitonic degrees of freedom\cite{Liu2014a, Dufferwiel2015,Lundt2017,Sidler2017,Dufferwiel2017}. Among these, the recently emerged exciton hybridization opens access to unexplored excitonic species \cite{Gerber2019,leisgang2020giant,lorchat2021excitons,Peimyoo2021,WilsonNature2021}, with a promise of enhanced interactions\cite{zhang2021van}. Here, we employ hybridized interlayer excitons (hIX) in bilayer MoS$_2$ \cite{Gerber2019,leisgang2020giant,lorchat2021excitons,Peimyoo2021} to achieve highly nonlinear excitonic and polaritonic effects. Such interlayer excitons possess an out-of-plane electric dipole \cite{leisgang2020giant} as well as an unusually large oscillator strength \cite{Gerber2019} allowing observation of dipolar polaritons (dipolaritons \cite{Cristofolini2012,Togan2018,Kyriienko2012}) in bilayers in optical microcavities. Compared to excitons and polaritons in MoS$_2$ monolayers, both hIX and dipolaritons exhibit $\approx 8$ times higher nonlinearity, which is further strongly enhanced when hIX and intralayer excitons, sharing the same valence band, are excited simultaneously. This gives rise to a highly nonlinear regime which we describe theoretically by introducing a concept of hole crowding. The presented insight into many-body interactions provides new tools for accessing few-polariton quantum correlations \cite{berloff2017realizing,delteil2019towards,kyriienko2020nonlinear}}.

Excitons in two-dimensional transition metal dichalcogenides (TMDs) have large oscillator strengths and binding energies \cite{Wang2018}, making them attractive as a platform for studies of strong light-matter coupling in optical microcavities \cite{Liu2014a, Dufferwiel2015,Lundt2017,Sidler2017}. A variety of polaritonic states have been realised using monolayers of MX$_2$ (M=Mo, W; X=S, Se) embedded in tunable \cite{Dufferwiel2015,Sidler2017,Dufferwiel2017,Lyons2021} and monolithic microcavities \cite{Gillard2021,gu2019room,lundt2019optical,zhang2021van,tan2020interacting}. 


\begin{figure}[t]%
\includegraphics[width=1\textwidth]{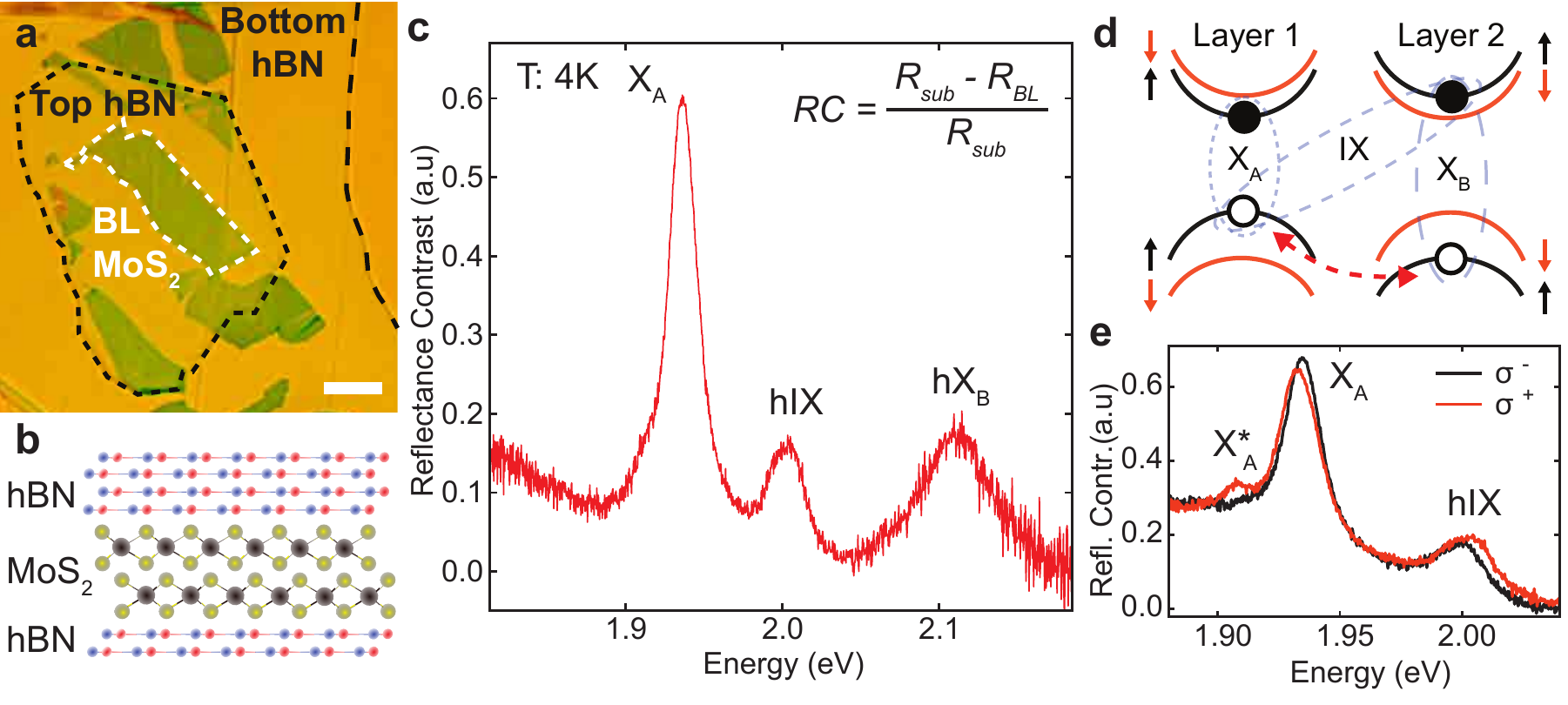}
\caption{\large \textbf{Homobilayer MoS$_2$ and its optical response. a,} Bright field microscope image of an encapsulated BL MoS$_2$ transferred on top of a DBR. Scale bar: 10 $\mu$m. \textbf{b,} Schematic side-view of the fabricated heterostructure comprising a BL MoS$_2$ sandwiched between few-layer hBN. \textbf{c,} Reflectance contrast (RC) spectrum of the sample measured at low temperature (4~K) showing three distinct absorption features at 1.937~eV, 2.004~eV and 2.113~eV for X$_A$, hIX and hX$_B$, respectively. The measured linewidths for X$_A$, hIX, and hX$_B$ are $20$, $23$ and $64$ meV, respectively. RC is calculated using the formula in the top-right corner of the graph. \textbf{d,} Sketch of the conduction and valence bands in two adjacent layers of MoS$_2$, displaying the allowed optical transitions of A and B direct intralayer excitons (X$_A$ and X$_B$) and interlayer excitons (IX) for spin-up states (black lines) at the K point in the bilayer momentum space. IX hybridizes with X$_B$ through the hole tunnelling between the two layers (red dashed arrow). At the K' point of the bilayer Brillouin zone, the same configuration applies for the states with the opposite spins. \textbf{e,} RC spectra of excitons in BL MoS$_2$ detected in two circular polarizations in an out-of-plane magnetic field of 8~T at $T$=4~K. Zeeman shifts of opposite signs are observed for X$_A$ and hIX. The absorption peak of the charged intralayer exciton (X$^*_A$) shows near unity circular polarization.}
\label{fig1}
\end{figure}

One of the central research themes in polaritonics is the study of nonlinear interactions leading to extremely rich phenomena such as Bose-Einstein condensation  \cite{Deng2010,Kasprzak2006}, polariton lasing \cite{Christopoulos2007,Bhattacharya2014} or optical parametric amplification \cite{AmoNat2009}. Polaritons formed from tightly bound neutral intralayer excitons in TMDs are not expected to show strong nonlinearity. However, pronounced nonlinear behavior was observed for trion polaritons \cite{Emmanuele2020,Lyons2021} and Rydberg polaritons \cite{Gu2021}. Enhanced nonlinearity can be achieved by employing excitonic states with a physically separated electron and hole, e.g. in adjacent atomic layers \cite{rivera2018interlayer} or quantum wells \cite{Butov2004,Kyriienko2012,Hubert2019,Cristofolini2012, Togan2018}. Such interlayer excitons have a large out-of-plane electric dipole moment, and thus can strongly mutually interact  \cite{butov2002macroscopically}. Typically, however, interlayer or 'spatially indirect' excitons possess low oscillator strength \cite{rivera2018interlayer,fox1991excitonic}. Thus, in order to strongly couple to cavity photons, hybridization with high-oscillator-strength intralayer excitons is required \cite{Cristofolini2012,Togan2018,Kyriienko2012,alexeev2019resonantly,zhang2021van}. 

An attractive approach for realization of dipolar excitons and polaritons is to employ the recently discovered exciton hybridization in MoS$_2$ bilayers \cite{Gerber2019,paradisanos2020controlling}. This approach  allows realization of uniform samples suitable for the observation of macroscopic many-body phenomena \cite{amo2009collective}. Interlayer excitons unique to bilayer MoS$_2$ possess a large oscillator strength, comparable to that of the intralayer exciton, arising from interlayer hybridization of valence band states, aided by a favourable orbital overlap and a relatively small spin-orbit splitting among semiconducting TMDs \cite{Gerber2019}. Such hybridized interlayer excitons (hIX) are highly tunable using out-of-plane electric field \cite{leisgang2020giant,lorchat2021excitons} and their valley degree of freedom persists up to room temperature \cite{Peimyoo2021}. 


Here we use hIXs in bilayer MoS$_2$ to realize highly nonlinear excitonic and dipolaritonic effects. We unravel a previously unexplored interaction
regime involving intra- and interlayer excitons stemming from the fermionic nature of the charge carriers in a valence band shared between different excitonic species. This regime, accessible using broadband excitation resonant with both hIX and intralayer exciton transitions, provides strong (up to 10 times) enhancement of the exciton nonlinearity, already enhanced by up to 8 times in MoS$_2$ bilayers compared with monolayers.  We support our experimental findings with microscopic theory, analysing the excitonic many-body physics and the cross-interactions and introducing the nonlinear mechanisms of the hole crowding.

\begin{figure}[t]
\centering
\includegraphics[width=0.8\textwidth]{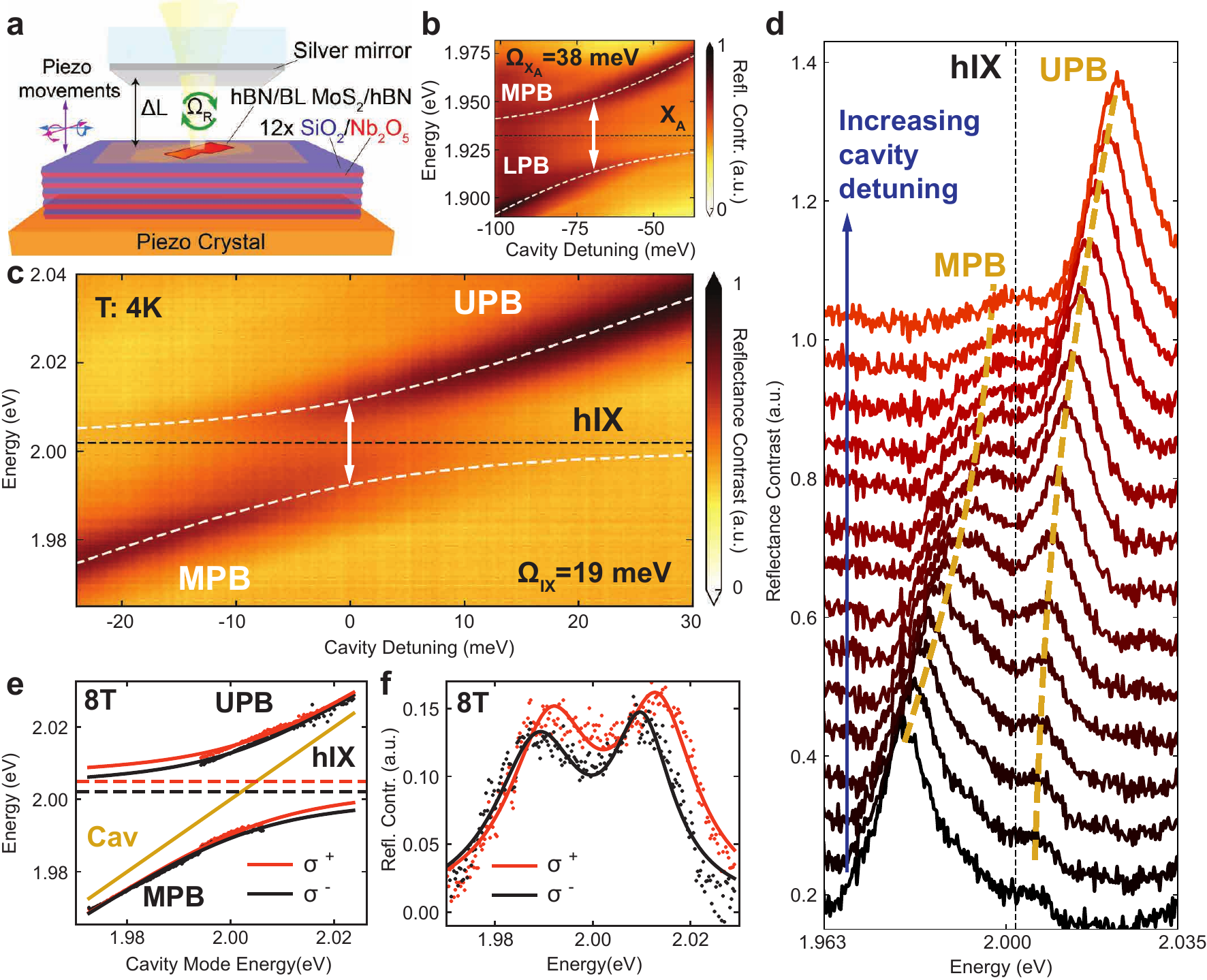}
\caption{\large {\bf Strong exciton-photon coupling in MoS$_2$ bilayers. a,} Schematics of the tunable open microcavity composed of a bottom DBR and a top semi-transparent silver mirror. 
{\bf b, c,} Low temperature (4K) RC spectra measured as a function of the cavity-exciton detuning ($\Delta = E_{\mathrm{cav}} - E_{\mathrm{exc}}$) for cavity scans across \textbf{b} X$_A$ and \textbf{c} hIX energies.  White dotted lines show the fitting obtained using the coupled-oscillator model providing the Rabi splittings $\Omega_{\mathrm{hiX}}=19$ meV and $\Omega_{\mathrm{X_A}}=38$ meV. \textbf{d,} RC spectra measured for the cavity-exciton detunings in the vicinity of the anticrossing between hIX and the cavity mode. {\bf e,} Dipolariton dispersion measured with circularly polarized detection for 8~T magnetic field. The orange and black solid curves are the coupled oscillator model fits for $\sigma^{+}$ and $\sigma^{-}$ detection, respectively. The positions of the Zeeman-split hIX peaks are shown by dashed lines. {\bf f,} $\sigma^{+}$ (orange) and $\sigma^{-}$ (black) RC spectra measured at 8~T at the hIX-cavity anticrossing. Fitting with two Lorentzians (solid lines) is shown.}\label{fig2}
\end{figure}


Our heterostructure samples consists of  a MoS$_2$ bilayer (BL) sandwiched between hBN and placed on a distributed Bragg reflector (DBR). Fig.~\ref{fig1}a shows a bright field microscope image of the encapsulated BL MoS$_2$. A sketch of the side view of the device is displayed in Fig.~\ref{fig1}b.  The reflectance contrast (RC) spectrum of the studied MoS$_2$ bilayer, displayed in Fig.~\ref{fig1}c, shows three peaks: the intralayer neutral excitons X$_A$ at at $1.937$~eV (see Fig.~\ref{fig1}d), hybridized interlayer exciton hIX at  $2.004$~eV and  hybridized B-exciton at $2.113$~eV. Due to the quantum tunnelling of holes, B-excitons hybridize with an interlayer exciton (IX) (Fig.~\ref{fig1}d), which is a direct transition in the bilayer momentum space \cite{Gerber2019}.  The ratio of the integrated intensities of X$_A$ and hIX is $4.5$. Based on these data, we estimate the electron-hole separation $d = 0.55$~nm (see details in Supplementary Note S1) in agreement with previous studies \cite{Peimyoo2021}. We further confirm the nature of the hIX states by placing the BL MoS$_2$ in magnetic field where the valley degeneracy is lifted (Fig. \ref{fig1}e). In agreement with recent studies \cite{lorchat2021excitons,Gong2013}, we measure a Zeeman splitting with an opposite sign and larger magnitude in hIX compared with X$_A$ (-3.5 versus 1.5 meV).


We study the strong coupling regime in a tunable planar microcavity (Fig.~\ref{fig2}a) formed by a silver mirror and a planar DBR \cite{Dufferwiel2015}. 
RC scans as a function of the cavity mode detuning $\Delta = E_{cav} - E_{exc}$, where $E_{cav}$ and $E_{exc}$ are the cavity mode and the corresponding exciton energy, respectively, are shown in Fig. \ref{fig2}c,d. Characteristic anticrossings of the cavity mode with X$_A$ and hIX are observed, resulting in lower, middle and upper polariton branches (LPB, MPB, and UPB, respectively). The extracted Rabi splittings  are $\Omega_{\mathrm{X_A}} = 38$~meV for X$_A$ and $\Omega_{\mathrm{hIX}} = 19$~meV for hIX (Supplementary Note S2). Fig.~\ref{fig2}d shows the RC spectra in the vicinity of the anticrossing with hIX, providing a more detailed view of the formation of the MPB and UPB.  The intensity of the polariton peaks is relatively low for the states with a high exciton fraction at positive (negative) cavity detunings for the MPB (UPB). As the Rabi splitting scales as a square root of the oscillator strength, the ratio $\Omega_{\mathrm{X_A}}/\Omega_{\mathrm{hIX}} = 2$ is in a good agreement with the RC data for integrated intensities of X$_A$ and hIX. From the Rabi splitting ratio we can estimate the tunneling constant $J$ leading to the exciton hybridization. The corresponding coefficient is $J = 48$~meV (see Supplementary Note S1 for details), matching the density functional theory predictions \cite{Gerber2019}. In polarization-resolved cavity scans in an out-of-plane magnetic field (Fig.~\ref{fig2}e,f), similarly to hIX behaviour, we observe opposite and larger Zeeman splitting for dipolaritons relative to the intralayer polaritons (see Supplementary Figure S4). Chiral dipolariton states are observed distinguished by their opposite circular polarization (Fig.~\ref{fig2}f).



\begin{figure}[t]
\centering
\includegraphics[width=1\textwidth]{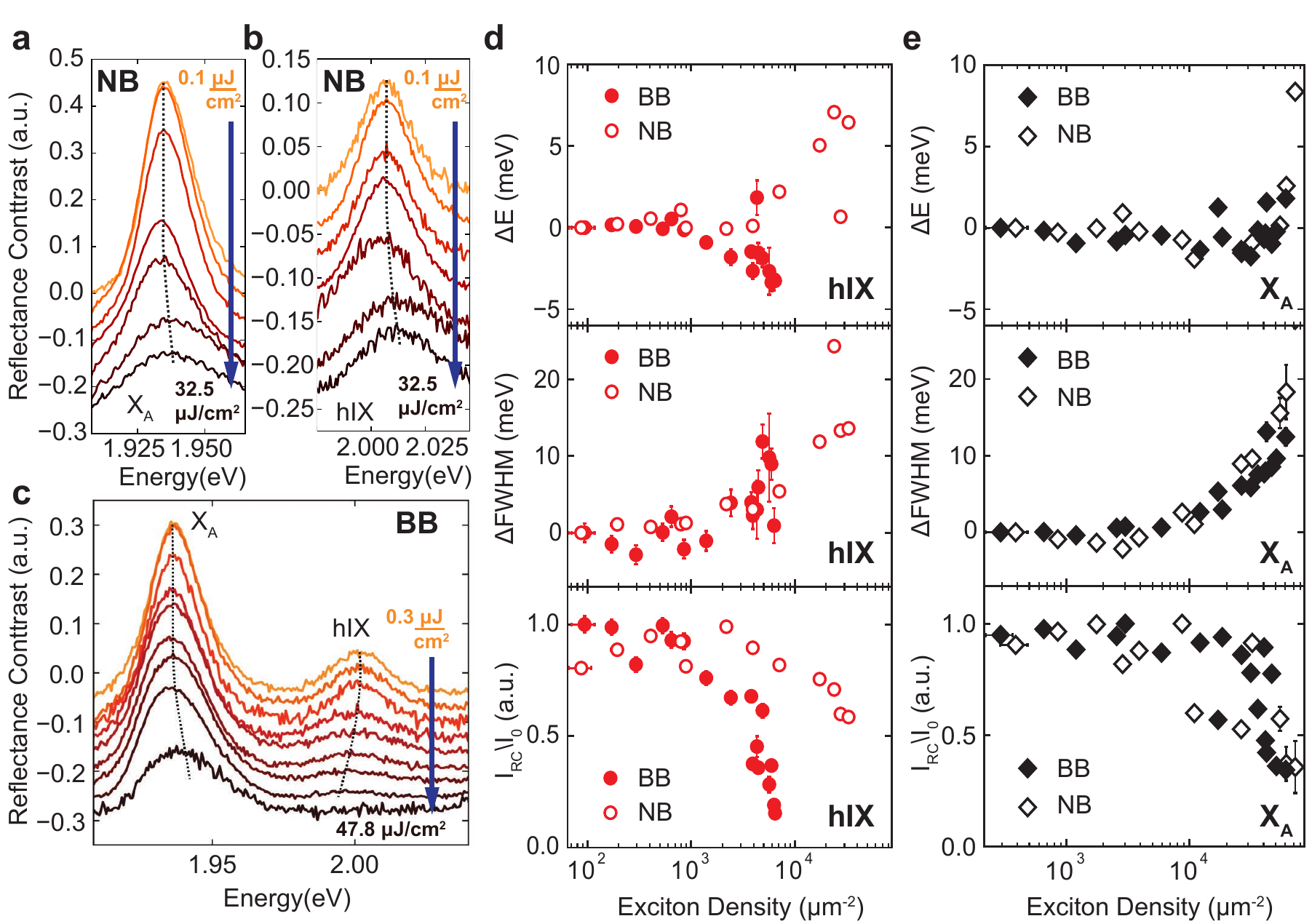}
\caption{\large {\bf Exciton nonlinearity in MoS$_2$ bilayers. a, b, c,} RC spectra measured with the NB  (FWHM=28nm) excitation for the X$_A$ (\textbf{a}) and hIX (\textbf{b}), and with the BB (FWHM=50nm) excitation (\textbf{c}) at different fluences. The dashed curves are guide for the eye. {\bf d, e,} The energy shift $\Delta E$ (top), linewidth variation $\Delta$FWHM (middle) and normalized integrated intensity (bottom) as a function of the exciton density for the hIX (d) and X$_A$ (e). Solid (open) symbols show the results for the BB (NB) excitation. For the normalized intensity we divide the intergrated intensity at each laser power by that at the maximum intensity.}\label{fig3}
\end{figure}

We investigate the nonlinear response of X$_A$ and hIX in the bare BL flake as a function of the laser power using both narrow band (NB, full-width at half maximum, FWHM=28nm) and broad band (BB, FWHM=50 nm) pulsed excitation (see Methods). Our resonant pump-probe experiments have confirmed that the lifetimes of the hIX and X$_A$ states are considerably longer than the pulse duration of $\approx150$ fs (Supplementary Note S3). Measured RC spectra are shown in Fig.~\ref{fig3}a,b for the NB and in Fig.~\ref{fig3}c for BB excitation. In the NB case, the excitation was tuned to excite either X$_A$ or hIX independently, while in the BB case, both resonances were excited simultaneously. 

As seen in Figs.~\ref{fig3}a,b both X$_A$ and hIX spectra behave similarly upon increasing the power of the NB excitation: a blueshift of several meV is observed, accompanied by the peak broadening and bleaching. For the BB excitation, however, a different nonlinear behaviour is observed as shown in Fig.~\ref{fig3}c: the broadening and complete suppression of the hIX peak is observed at much lower powers, accompanied by a redshift. This is in contrast to X$_A$, whose behaviour is similar under the two excitation regimes. 

The resulting energy shifts, peak linewidths and intensities are shown in Fig.~\ref{fig3}d,e as a function of the exciton density (see details in Supplementary Note S4 and S6). Fig.~\ref{fig3}d quantifies the trends observed in Figs.~\ref{fig3}a,b showing for the BB excitation an abrupt bleaching of the hIX peak above the hIX density $5\times 10^3~\mu$m$^{-2}$ accompanied by a redshift of $\approx$ 4 meV and a 12 meV broadening. For the NB case, a similar decrease in peak intensity is observed only around $4\times 10^4~\mu$m$^{-2}$, accompanied with a peak blueshift of $\approx$ 7 meV and a broadening exceeding 15 meV.  In Fig.~\ref{fig3}e, however, it is apparent that the observed behaviour under the two excitation regimes is similar for X$_A$. A similar blueshift, broadening and saturation are observed at slightly higher densities compared to the hIX under the NB excitation (Supplementary Note S5). We also find that due to the increased excitonic Bohr radius, the onset of the nonlinear behaviour for X$_A$ in bilayers occurs at a lower exciton density than for X$_A$ in monolayers (Supplementary Note S7). 

\begin{figure}[t]
    \centering
    \includegraphics[width = 0.6 \linewidth]{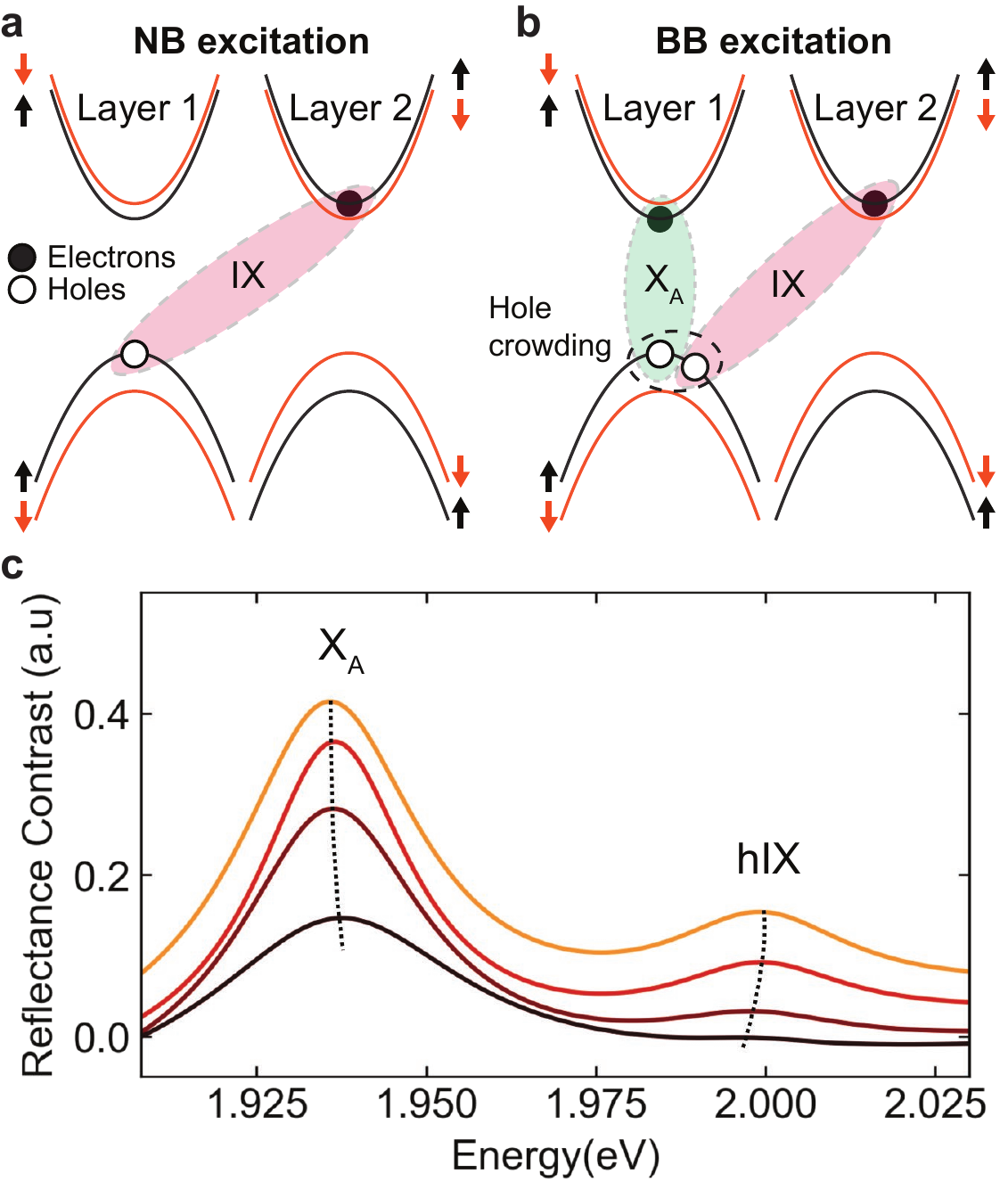}
     \caption{\large \textbf{Theoretical model for nonlinear optical response in MoS$_2$ bilayers.} \textbf{a, b,} Schematic diagram showing exciton generation under the NB (a) and BB (b) excitation. In (a) only generation of hIX is shown. In (b), the holes of the two excitonic species share the same valence band. \textbf{c,} Theoretically calculated absorption spectra for the BB excitation case (see Supplementary Note S8), providing qualitative agreement with the experiment. The dashed black curves are guides for the eye.}
    \label{fig:theory-figure}
\end{figure}

We develop a microscopic model to describe the contrasting phenomena under the NB and BB excitation. Under the NB excitation, either X$_A$ or hIX excitons are created as sketched in Fig.~\ref{fig:theory-figure}a. In this case, nonlinearity arises from Coulomb exciton-exciton interactions causing the blueshift and dephasing \cite{Erkensten2021}. For simplicity, in the main text we will use a Coulomb potential $V_{Coul}$ combining the exchange and direct terms further detailed in Supplementary Note S8. We confirm (see Supplementary Note S8) that for the intralayer exciton-exciton interaction (X$_A$-X$_A$) the dominant nonlinear contribution comes from the Coulomb exchange processes, as in the monolayer case \cite{Erkensten2021,Shahnazaryan2017}, while for the hIX-hIX scattering the dominant contribution is from the direct Coulomb (dipole-dipole) interaction terms \cite{Kyriienko2012}. For both X$_A$ and hIX, the Coulomb interaction is repulsive, and thus leads to the experimentally observed blueshifts. We find that for the modest electron-hole separation $d=0.55$~nm in the bilayer, $V_{Coul}$ is overall 2.3 times stronger for hIX compared with X$_A$. 


Analysing the shapes of the reflectance spectra in the NB case, we note that they depend on the rates of radiative ($\Gamma_{\mathrm{R}}$) and non-radiative ($\Gamma_{\mathrm{NR}}$) processes. The area under RC curves is described by the ratio $\Gamma_{\mathrm{R}} / (\Gamma_{\mathrm{R}} + \Gamma_{\mathrm{NR}})$. This ratio changes under the increased excitation if the rates depend on the exciton densities. Specifically, we account for the scattering-induced non-radiative processes that microscopically scale as $\Gamma_{\mathrm{NR}} \propto \vert V_{\mathrm{Coul}} \vert^2 n$, i.e. depend on the absolute value of the combined matrix elements for the Coulomb interactions and the exciton density $n$ \cite{Erkensten2021}. This process allows reproducing the RC behaviour and bleaching at increasing pump intensity. Moreover, it explains stronger nonlinearity for X$_A$ in bilayers compared to monolayers. 
Namely, the scattering scales with the exciton Bohr radius, $V_{\mathrm{Coul}} \propto \alpha$, which is larger in the bilayers due to the enhanced screening (Supplementary Note S8).

In the BB case,  both X$_A$ and hIX excitons are generated simultaneously, and together with intraspecies scattering (X$_A$-X$_A$ and hIX-hIX), interspecies scattering (X$_A$-hIX) occurs, similarly to the direct-indirect exciton Coulomb scattering in double quantum wells \cite{Kristinsson2013}. Since X$_A$ and hIX are formed by the holes from the same valence band (Fig.~\ref{fig:theory-figure}a), an additional contribution arises from the phase space filling, i.e. the commutation relations for the excitons (composite bosons) start to deviate from the ideal weak-density limit once more particles are created \cite{Combescot2008}. For particles of the same flavour, the phase space filling enables nonlinear saturation effects in the strong coupling regime, similar to polariton saturation observed in \cite{Emmanuele2020}. However, in the presence of several exciton species, we reveal a distinct phase space filling mechanism which we term the {\it hole crowding}. Crucially, we observe that the commutator of the X$_A$ annihilation operator ($\hat{X}$) and hIX creation operator ($\hat{I}^\dagger$) is non-zero, $[\hat{X}(\mathbf{p}), \hat{I}^\dagger(\mathbf{q})] = - \hat{B}_{\mathbf{p}, \mathbf{q}}$. Here $\mathbf{p}$, $\mathbf{q}$ are exciton momenta and $\hat{B}_{\mathbf{p}, \mathbf{q}}$ is an operator denoting the deviation from the ideal commuting case ($\hat{B}_{\mathbf{p}, \mathbf{q}} = 0$) of distinct bosons where holes do not compete for the valence band space. 

This statistical property of modes that share a hole has profound consequences for the nonlinear response. Namely, the total energy is evaluated as an expectation value over a many-body state with both X$_A$ and hIX excitons, $\vert N_{\mathrm{X}}, N_{\mathrm{hIX}} \rangle := (\prod_{\mathbf{p}}^{N_{\mathrm{X}}} \hat{X}^\dagger) (\prod_{\mathbf{q}}^{N_{\mathrm{hIX}}}\hat{I}^\dagger) \vert \Omega_{max} \rangle $, where $N_{\mathrm{X}}$ and $N_{\mathrm{hIX}}$ particles are created from the ground state $\vert \Omega_{max} \rangle$. If the excitonic modes are independent, the contributions from X$_A$ and hIX simply add up. However, the hole coexistence in the valence band induces the excitonic interspecies scattering. The phase space filling combined with the Coulomb energy correction leads to a negative nonlinear energy contribution.  This nonlinear term scales as $\Delta E_{\mathrm{hIX}} = - \eta \sqrt{n_{\mathrm{X}} n_{\mathrm{hIX}}}$, where $\eta > 0$ is a coefficient defined by the Coulomb energy and Bohr radii and $n_{\mathrm{X,hIX}}$ are the exciton densities (see Supplementary Note S9).  This nonlinearity also modifies the non-radiative processes leading to substantial broadening for the hIX states. 

According to this analysis, the effect of the BB excitation should be most pronounced for hIX. In addition to the possible hIX-hIX scattering (similar to that occurring under the NB excitation), much stronger X$_A$ absorption leads to the phase space filling in the valence band. Such hole crowding introduces additional scattering channels for hIX and leads to its RC spectra bleaching at lower hIX exciton densities. On the other hand, as only relatively small hIX densities can be generated, both the NB and BB excitation cases should produce similar results for X$_A$. Using the estimated nonlinear coefficients caused by the hole crowding, we model the RC in the BB regime and qualitatively reproduce the strong bleaching and redshift for hIX at the increased density. 



\begin{figure}[h]
\centering
\includegraphics[width=1\textwidth]{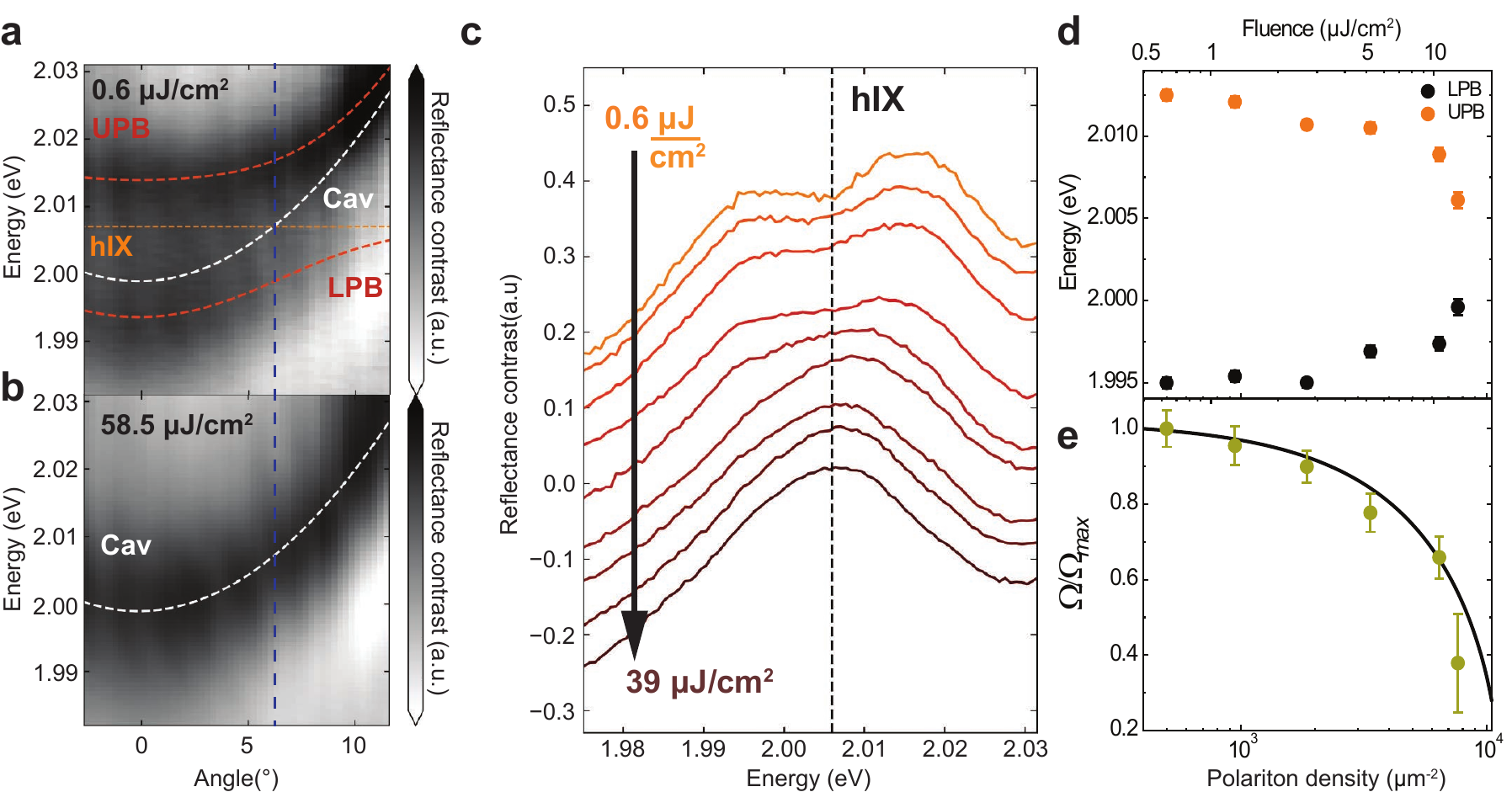}
\caption{{\bf Nonlinear behaviour of dipolaritons. a, b,} Reflectance contrast spectra measured at different laser fluences for the MoS$_2$ bilayer placed in a monolithic cavity. (a) The low fluence case (0.6 $\mu$J cm$^{-2}$). A clear anticrossing at 6.5\textdegree is observed. Dashed red lines show the results of the fitting using a coupled oscillator model, with two polariton branches LPB and UPB formed. White and orange lines show the energies of the uncoupled cavity mode and hIX state, respectively. The vertical line marks the anticrossing angle. (b) The high fluence case (58.5 $\mu$J cm$^{-2}$). A complete collapse of the strong coupling regime is observed, with the disappearance of the anticrossing and transition into the weak coupling regime. {\bf c,} RC spectra measured at the anticrossing at 6.5{\textdegree}  as a function of the laser fluence.   {\bf d,} Measured UPB and LPB peak energies at 6.5{\textdegree}  as a function of the laser fluence (see top axis) and the corresponding polariton density (bottom axis). {\bf e,} Symbols show the Rabi splittings normalized by the Rabi splitting measured at the lowest power ($\Omega$/$\Omega_{max}$) as deduced from {\bf d,}. The line shows the fitting using our theoretical model (Supplementary Note S8).}\label{fig5}
\end{figure}

We investigate nonlinear properties of dipolar polaritons in a monolithic (fixed-length) cavity created by a silver mirror on top of a PMMA spacer (245 nm thick) covering the hBN-encapsulated MoS$_2$ homobilayer placed on the DBR. The cavity mode energy can be tuned by varying the angle of observation (0 degrees corresponds to normal incidence). We use a microscopy setup optimized for Fourier-plane imaging, thus allowing simultaneous detection of reflectivity spectra in a range of angles as shown in Fig.\ref{fig5}(a) displaying the measured polariton dispersion.  In this experiment, the cavity mode is tuned around hIX and only two polariton branches LPB and UPB are observed at low fluence of 0.6 $\mu$J cm$^{-2}$ with a characteristic Rabi splitting of 17.5 meV. In Fig.\ref{fig5}(b), at an increased fluence of 58.5 $\mu$J cm$^{-2}$, only a weakly coupled cavity mode is visible.
 
Fig.\ref{fig5}(c) shows RC spectra taken at $\sim6.5${\textdegree}  around the anticrossing at different laser fluences. The collapse of the two polariton  peaks into one peak signifying the transition to the weak coupling regime is observed above 25 $\mu$J cm$^{-2} $.  The LPB and UPB energies extracted using the coupled oscillator model (Supplementary Figure S5) are shown in Fig.\ref{fig5}(d). As the polariton density is increased, the LPB and UPB approach each other almost symmetrically, converging to the exciton energy. The corresponding normalized Rabi splitting ($\Omega/\Omega_{max}$, where $\Omega_{max}$ is measured at low fluence) are shown in Fig.\ref{fig5}(d,e) as a function of the total polariton density. 

In this experiment, the cavity mode is considerably above the X$_A$ energy, which therefore is not coupled to the cavity. Hence, the extracted Rabi splittings are fitted with a theoretically predicted trend of $\Omega$ for the NB excitation regime (Supplementary Note S8). A nonlinear polariton coefficient $\beta = 0.86~\mu$eV$\mu$m$^{2}$ is extracted by differentiating the fitted function with respect to the polariton density. Comparing our results to X$_A$ intralayer-exciton-polaritons in monolayers in similar cavities \cite{zhang2021van}, we observe that the nonlinearity coefficient for dipolar interlayer polaritons is about an order of magnitude larger. This is in a good agreement with the theoretically predicted intrinsic nonlinearity of hybridized interlayer polaritons (Supplementary Note S8), and with our experimental data comparing hIX and monolayer X$_A$ outside the cavity (Supplementary Note S7). 

In summary, we report the nonlinear exciton and exciton-polariton behaviour in MoS$_2$ homobilayers, a unique system where hybridized interlayer exciton states can be realized having a large oscillator strength. We find that nonlinearity in MoS$_2$ bilayers can be enhanced when both the intralayer and interlayer states are excited simultaneously, the regime that qualitatively changes the exciton-exciton interaction through the hole crowiding effect introduced theoretically in our work. In this broad-band excitation regime, the bleaching of the hIX absorption occurs at 8 times lower hIX densities compared to the case when the interlayer excitons are generated on their own. In addition to this, we find that the dipolar nature of hIX states in MoS$_2$ homobilayers already results in 10 times stronger nonlinearity compared with the intralayer excitons in MoS$_2$ monolayers. Thus, we report on an overall enhancement of the nonlinearity by nearly two orders of magnitude. Thanks to the large oscillator strength, hIX can enter the strong coupling regime in MoS$_2$ bilayers placed in microcavities, as realized in our work. Similarly to hIX states themselves, dipolar polaritons also show 10 times stronger nonlinearity compared with exciton-polaritons in MoS$_2$ monolayers. We expect that in microcavities where the cavity mode is coupled to both hIX and X$_A$ in MoS$_2$ bilayers, and the excitation similar to the broad-band regime can thus be realized, the nonlinear polariton coefficient will be dramatically enhanced owing to the hole crowding effect, allowing highly nonlinear polariton system to be realized. We thus predict that MoS$_2$ bilayers will be an attractive platform for realization of quantum-correlated polaritons with applications in polariton logic networks \cite{berloff2017realizing} and polariton blockade \cite{delteil2019towards,kyriienko2020nonlinear}.

\section{Methods}\label{sec4}
The hBN/MoS$_2$/hBN heterostructures were assembled using a PDMS polymer stamp method. The PMMA spacer for the monolithic cavity was deposited using a spin-coating technique, while a silver mirror of 45 nm was thermally evaporated on top of it.

Broad-band excitation was used to measure the reflectance contrast (RC) spectra of the devices at cryogenic temperatures (4K), defined as $\mathrm{RC} = (R_{\mathrm{sub}}-R_{\mathrm{BL}})/R_{\mathrm{sub}}$, where $R_{\mathrm{sub}}$ and $R_{\mathrm{BL}}$ are the substrate and MoS$_2$ bilayer reflectivity, respectively. 
For the magnetic field studies the same RC measurements were performed using unpolarized light in excitation with polarizers, $\lambda/4$ polarizers and $\lambda/2$ waveplates in collection, to resolve $\sigma^{+}$ and $\sigma^{-}$ polarization.  The low temperature measurements using the tunable cavity were carried out in a liquid helium bath cryostat (T=4.2K) equipped with a superconducting magnet and free beam optical access. We used a white light LED as a source. RC spectra were measured at each $\Delta L$ and are integrated over the angles within 5 degrees from normal incidence.  The RC spectra measured in the cavity are fitted using Lorenzians. The peak positions are then used to fit to a coupled oscillator model, producing the Rabi splitting and the exciton and cavity mode energies. 

The measurements on the monolithic cavity were performed in a closed loop helium flow cryostat (T=6K). For the power-dependent RC experiments, we used supercontinuum radiation produced by 100 fs Ti:Sapphire laser pulses at 2 kHz repetition rate at 1.55 eV propagating through a thin sapphire crystal. The supercontinuun radiation was then filtered to produce the desired narrow-band excitation.

All the exciton and polariton densities were calculated following the procedure introduced by L. Zhang et al. \cite{zhang2021van}, taking into account the spectral overlap of the spectrum of the excitation laser and the investigated exciton peak (see further details in Supplementary Note S4).

\section{Acknowledgements}

CL, AG, TPL, SR and AIT acknowledge financial support of the European Graphene Flagship Project under grant agreement 881603 and EPSRC grants EP/V006975/1,  EP/V026496/1, EP/V034804/1 and EP/S030751/1. TPL acknowledges financial support from the EPSRC Doctoral Prize Fellowship scheme. CT, SDC and GC acknowledge support by the European Union Horizon 2020 Programme under Grant Agreement 881603 Graphene Core 3. AG and GC acknowledge support by the European Union Marie Sklodowska-Curie Actions project ENOSIS H2020-MSCA-IF-2020-101029644. PC, RJ and DGL thank EPSRC Programme Grant ‘Hybrid Polaritonics’ (EP/M025330/1).

\section{Author contributions}

CL and SR fabricated and characterized hBN-encapsulated MoS$_2$ samples. KW and TT synthesized the high quality hBN. CL and AG designed the microcavity samples. PC, RJ, DGL fabricated the microcavity samples. CL, AG, CT, TL and SDC carried out optical spectroscopy experiments. SC and OK developed theory. AG calculated polariton densities. CL and AG analyzed the data with contribution from AIT, TL, SC, OK, CT, SDC and GC. CL, AG, SC, OK and AIT wrote the manuscript with contribution from all other co-authors. AIT, OK, DGL, GC managed various aspects of the project. AIT supervised the project.

\pagebreak

\widetext
\renewcommand{\thesection}{Supplementary Note S\arabic{section}:}  
\renewcommand{\thetable}{S\arabic{table}}  
\renewcommand{\thefigure}{\textbf{S\arabic{figure}}}
\renewcommand{\theequation}{S\arabic{equation}}
\renewcommand{\figurename}{\textbf{Supplementary Figure}}

\setcounter{equation}{0}
\setcounter{figure}{0}
\setcounter{table}{0}
\setcounter{page}{1}
\begin{center}
\textbf{\Large Supplementary Information: Nonlinear Interactions of Dipolar Excitons and Polaritons in MoS\textsubscript{2} Bilayers}

\Large{Charalambos Louca,$^{1,\star}$ Armando Genco,$^{2, \dagger}$ Salvatore Chiavazzo,$^{3}$ Thomas
P. Lyons,$^{1, 4}$ Sam Randerson,$^{1}$ Chiara Trovatello,$^{2}$ Peter Claronino,$^{1}$ Rahul
Jayaprakash,$^{1}$ Kenji Watanabe,$^{5}$ Takashi Taniguchi,$^{5}$ Stefano Dal Conte,$^{2}$ David G.
Lidzey,$^{1}$ Giulio Cerullo,$^{2}$ Oleksandr Kyriienko,$^{3}$ and Alexander I. Tartakovskii$^{1, \ddag}$\\}
\textit{\small$^{1}$Department of Physics and Astronomy, The University of Sheffield, Sheffield S3 7RH, UK\\ \vspace{1ex}
$^{2}$Dipartimento di Fisica, Politecnico di Milano, Piazza Leonardo da Vinci, 32, Milano, 20133, Italy\\ \vspace{1ex}
$^{3}$Department of Physics, University of Exeter, Stocker Road, Exeter, EX4 4PY, UK\\ \vspace{1ex}
$^{4}$RIKEN Center for Emergent Matter Science, Wako, Saitama, 351-0198, Japan\\
$^{5}$Advanced Materials Laboratory, National Institute for Materials Science, 1-1 Namiki, Tsukuba, 305-0044, Japan}
\end{center}
\vspace{1cm}

\section{Supplementary Note S1: Theoretical estimate of exciton properties --- energy and hybridisation}\label{sec:exciton_properties}

In this section of Supplemental Materials we describe details of theoretical description of MoS\textsubscript{2} homobilayers. Properties of excitons in a bilayer system have been discussed in the main text, and we support them by modelling. The optical response of the system is characterised by the response of three different species of quasi-particles, namely X\textsubscript{A}, hIX and hX\textsubscript{B} {\color{blue}(see main text)}. Here we provide an intuitive picture of the homobilayer physics and estimate the system parameters. In particular, we estimate exciton Bohr radii and a hole tunnelling rate as relevant parameters when studying nonlinear properties. 
\begin{figure}[h!]
    \centering
    \includegraphics[width=0.6\linewidth]{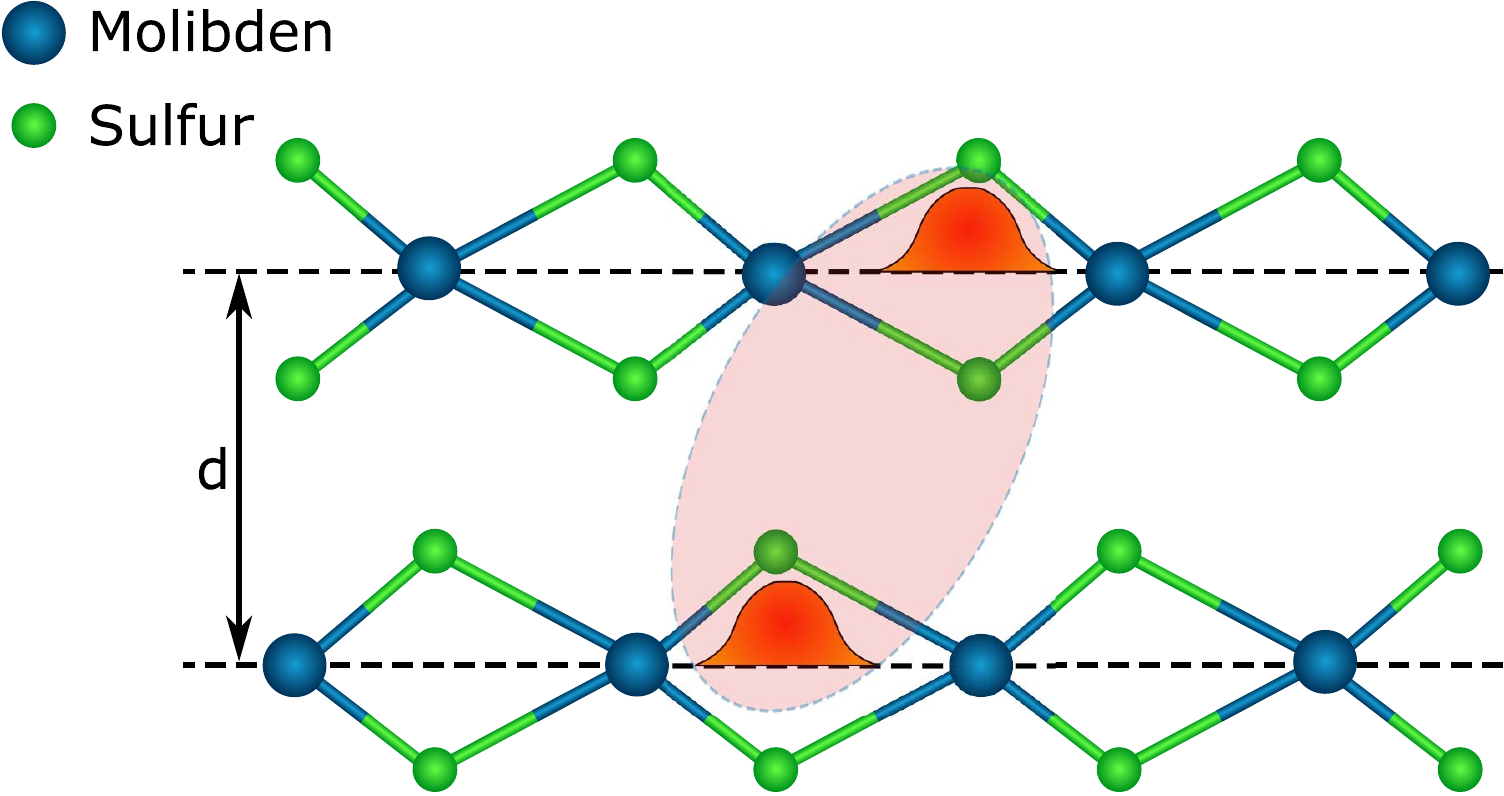}
    \caption{\textbf{Side view of 2H-stacked MoS\textsubscript{2} bilayer.} Blue spheres are Mo (molybdenum) atoms and green are S (sulphide) atoms. We picture a hole wave function in each layer, constructing a hybrid state through tunneling process (hole delocalisation). This allows for hybrid hX\textsubscript{B} and hIX excitons. We define the distance between the layer as the distance between particle centres of charge. In absence of external fields, the centre of charge is located in between the S planes \cite{Cappelluti2013}.}
    \label{fig:latticeProfile}
\end{figure}

\begin{figure}[ht!]
    \centering
    \includegraphics[width=\linewidth]{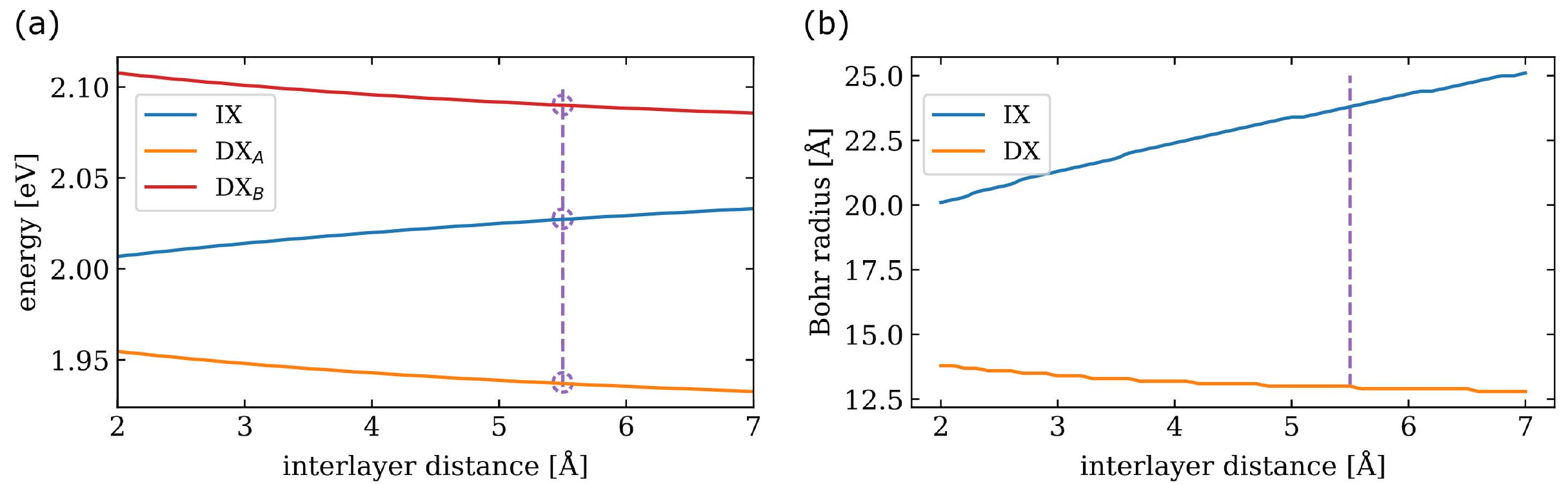}
    \caption{\textbf{Evolution of particle properties with interlayer distance. a)} Energy of quasiparticle modes with distance. The blue line is the hIX mode, and the orange and red respectively are X\textsubscript{A} and hX\textsubscript{B}. The redshift of direct modes with the increase of interlayer distance is due to reduced screening effects, as characteristic of Keldysh-Rytova potential in Eq. \eqref{eqs:Keldysh}. On the contrary, the hIX mode sees a blueshift that is due to the lower attraction of particles located in separate layers. The purple dashed line is the distance to match with the experimental data. \textbf{b)} Dependence of Bohr radius with interlayer distance. Here we only report one direct exciton as both A and B excitons have the same behavior. The orange one is the evolution for direct excitons, and the blue line is for indirect excitons. Both plots (a) and (b) show similiar increase (or increase) with distance as the screening mechanism affects the considered parameters the same way. This analysis reveals the hIX Bohr radius to be roughly twice as much as the X one.}
    \label{fig:bindings}
\end{figure}

We consider a MoS\textsubscript{2} homobilayer system with 2H stacking (see Fig.~\ref{fig:latticeProfile}). This is comprised of two parallel layers of MoS\textsubscript{2}, with centres located at a distance $d$ from each other (charge separation distance). The physics of bilayers is defined by properties of electrons and holes that interact through the Keldysh-Rytova potential \cite{Cudazzo2011, Berkelbach2013,Chernikov2014}, being different for in-plane and out-of-plane interaction \cite{Danovich2018}. Within the $k \cdot p$ framework, electrons and holes are treated as particles with effective mass provided by a band dispersion. In MoS\textsubscript{2} the typical values for effective masses are $0.46$~m\textsubscript{e} for conduction bands and $0.56$~m\textsubscript{e} for the valence bands (with m\textsubscript{e} being the electron mass) \cite{Gerber2019, Kormanyos2015}. The attractive Kledysh-Rytova potential has a different form depending on the relative position between particles. We call $V^{\mathrm{intra}}_{\mathrm{KR}}$ the attractive potential of particles being in the same layer, and $V^{\mathrm{inter}}_{\mathrm{KR}}$ the attractive potential of particles being in separate layers. In momentum space the different potentials read as
\begin{subequations}\label{eqs:Keldysh}
\begin{align}
    V^{\mathrm{intra}}_{\mathrm{KR}}(\mathbf q) =&
    - \frac{e^2}{4 \pi \epsilon \epsilon_{\mathrm{0}}} \frac{1 + (r_{\mathrm{0}} q/\epsilon) (1 - \exp(-2 q d))}{(1 + r_{\mathrm{0}} q/\epsilon)^2 - (r_{\mathrm{0}} q / \epsilon)^2 \exp(- 2 q d)},\\
    V^{\mathrm{inter}}_{\mathrm{KR}}(\mathbf q) =& - \frac{e^2}{4 \pi \epsilon \epsilon_{\mathrm{0}}} \frac{\exp(-q d)}{(1 + r_{\mathrm{0}} q/\epsilon)^2 - (r_{\mathrm{0}} q / \epsilon)^2 \exp(- 2 q d)},
\end{align}
\end{subequations}
where $e$ is the electron charge, $\epsilon_{\mathrm{0}}$ is the vacuum permittivity, $\epsilon$ in an average environment permittivity, $r_{\mathrm{0}}$ is a screening length (defined as for monolayers), and $\mathbf{q}$ is an exchanged particle momentum \cite{Danovich2018}. We compute a binding energy of an exciton bound state by assuming the Keldysh-Rytova attractive potential and free particle dispersion defined by the effective masses. To provide a simple understanding of the system, we approach the problem using an ansatz wavefunction $\phi(\rho) = \sqrt{2 / \pi \alpha^2} \exp(- \rho / \alpha)$, with $\rho$ being the in-plane projection of electron-hole distance and $\alpha$ the exciton Bohr radius. This describes well an internal structure of an exciton, and gives the information about its shape, collected in the Bohr radius. In Fourier space, the function $\phi(q)$ reads
\begin{equation}\label{eq:ansatz}
    \phi(q) = \sqrt{\frac{2}{\pi}} \frac{\alpha}{(1 + q^2 \alpha^2)^{3/2}}.
\end{equation}
With the given wave function [Eq.~\eqref{eq:ansatz}], and the given potentials [Eq.~\eqref{eqs:Keldysh}], we can find the binding energy and the Bohr radius of the excitons by minimizing the energy of the system. This procedure is performed in the range of possible interlayer separation $d$. Fig.~\ref{fig:bindings}(a) shows the energy change with the separation. The blue curve corresponds to the hIX mode, while orange and red curves correspond to X\textsubscript{A} and hX\textsubscript{B}, respectively. We considered a band-gap of $2.12$~eV, and spin-orbit splitting of $13$~meV for conduction band and $150$~meV for valence band, as suggested by the ab initial calculations \cite{Kormanyos2015}.
Here we estimate the interlayer distance $d$ thanks to experimental knowledge of energy distance between exciton energy modes. Note the peak of X\textsubscript{A} mode is not shifted by any tunneling, while X\textsubscript{B} and IX are coupled through tunneling constant $J$ \cite{Alexeev2019}. That is, the theoretical energy distance $\Delta_{\mathrm 0}$ between the two modes has to be fixed to match the experimental result $\Delta = 109$ meV. With considering two coupled harmonic oscillators, we find the corrected energy shift to be
\begin{equation}
    \Delta = \sqrt{\Delta_{\mathrm{0}}^2 + 4 J^2}.
\end{equation}
By matching with the experimental data for the energy distance between modes, we extract $d = 5.5$~\r{A} and estimate the effective tunneling rate to be $J = 45$~meV. 
As a consequence, $21 \%$ of X\textsubscript{B} oscillator strength is transferred to IX mode, in agreement with previous observations \cite{Gerber2019}. Finally, Fig. \ref{fig:bindings}(b) shows the evolution of particle Bohr radii with the interlayer distance. We respectively call the Bohr radius of direct and indirect exciton $\alpha_{\mathrm{D}}$ and $\alpha_{\mathrm{I}}$. The blue curve is the Bohr radius of hIX, while the orange one described the X modes. With X\textsubscript{A} and hX\textsubscript{B} being very similar, we describe both with one orange curve. The energy separation is provided only by the spin-orbit splitting. Typical values of $\alpha_{\mathrm{D}}$ are approximately $1$~nm, with $\alpha_{\mathrm{I}}$ being approximately $2$~nm. Note the opposite behavior of $\alpha_{\mathrm{D}}$ and $\alpha_{\mathrm{I}}$ with distance. hIX Bohr radius grows with distance due to the reduced attraction between particles in separate layers. On the contrary, by increasing the interlayer distance, we see the reduced screening for particles in the same layer, resulting in a decrease of the Bohr radius and consequent increase of the binding energy.

\section{Supplementary Note S2: Coupled oscillator model}\label{CO}

A full picture of our system, corresponding to MoS\textsubscript{2} homobilayer, has to take into account 4 different modes coupling to each other: only IX and X\textsubscript{B} hybridise through a tunneling parameter, while X\textsubscript{A} and X\textsubscript{B} can couple with the cavity due to their high oscillator strength. We can then simplify this picture by rewriting the IX and X\textsubscript{B} states in terms of the new basis of hybridised modes, hIX and hX\textsubscript{B}, as defined in the main text, all of them now capable of a coupling with the cavity mode. The corresponding Hamiltonian reads
\begin{equation}
H=\left(\begin{array}{cccc}{E}_{\mathrm{c}} & \Omega_{\mathrm{X_{A}}} & \Omega_{\mathrm{hIX}} & \Omega_{\mathrm{hX_{B}}} \\ \Omega_{\mathrm{X_{A}}}  & E_{\mathrm{X_{A}}} & 0 & 0\\ \Omega_{\mathrm{hIX}} & 0 & E_{\mathrm{hIX}}& 0 \\ \Omega_{\mathrm{hX_{B}}} & 0 & 0 & E_{\mathrm{hX_{B}}}\end{array}\right),
\end{equation}
where $E_{\mathrm{c}}$ in the energy of the cavity mode, and $E_{\mathrm{X_{A}}}$, $E_{\mathrm{hIX}}$, $E_{\mathrm{hX_{B}}}$ denote energies of the respective excitonic modes. Here, $\Omega_{\mathrm{X_{A}}}$, $\Omega_{\mathrm{hIX}}$, and $\Omega_{\mathrm{hX_{B}}}$ are corresponding matrix elements for light-matter coupling (Rabi splittings).

\begin{figure}[h]
\centering
\includegraphics[width=1\textwidth]{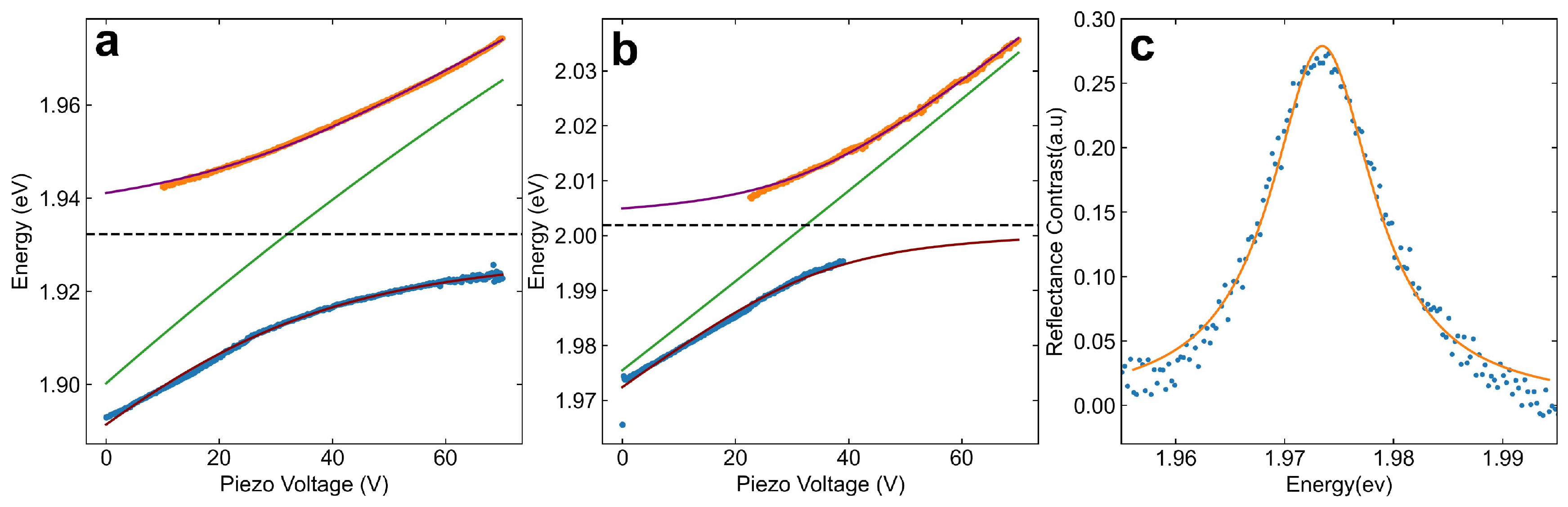}
\caption{\textbf{Coupled oscillator model fits. a), b)} Orange and blue dots represent the extracted peak energies of individual spectra near the a) X\textsubscript{A} and  b) hIX energies. The solid purple (UPB) and red (LPB) curves are the solutions to the fitted coupled oscillator Hamiltonians, and the green solid line is the extracted cavity mode energy as a function of voltage. \textbf{c)} shows a spectrum of the uncoupled cavity mode. A Lorenzian peak (orange) is fitted to the data (blue dots) and the linewidth is extracted to be equal to 11 meV, confirming that the condition for strong coupling regime is met, $\Omega^2 > ({\gamma_{c}}^2+{\gamma_{x}}^2)/2$ as stated by \cite{Savona1995}.  }\label{SI_CO}
\end{figure}

Due to the large energy separation between the resonances, each splitting can be fitted to a two level oscillator model independently. In our case, the spectra from the open cavity scans at piezo voltages close to resonant anticrossings between the cavity and either X\textsubscript{A} or hIX, were fitted with Lorenzian functions. The results were then fitted to the respective Hamiltonians of two coupled oscillators, such that  $H_{\mathrm{A}}=\left(\begin{array}{cc}{E}_{\mathrm{c}} & \Omega_{\mathrm{X_{A}}} \\ \Omega_{\mathrm{X_{A}}}  & E_{\mathrm{X_{A}}} \end{array}\right)$ and $H_{\mathrm{h_{IX}}}=\left(\begin{array}{cc}{E}_{\mathrm{c}} & \Omega_{\mathrm{hIX}} \\ \Omega_{\mathrm{hIX}}  & E_{\mathrm{hIX}} \end{array}\right)$ and the values of Rabi splittings and resonant energies were extracted. The results of the fit are shown in Fig. \ref{SI_CO}

\begin{figure}[h]
\centering
\includegraphics[width=0.8\textwidth]{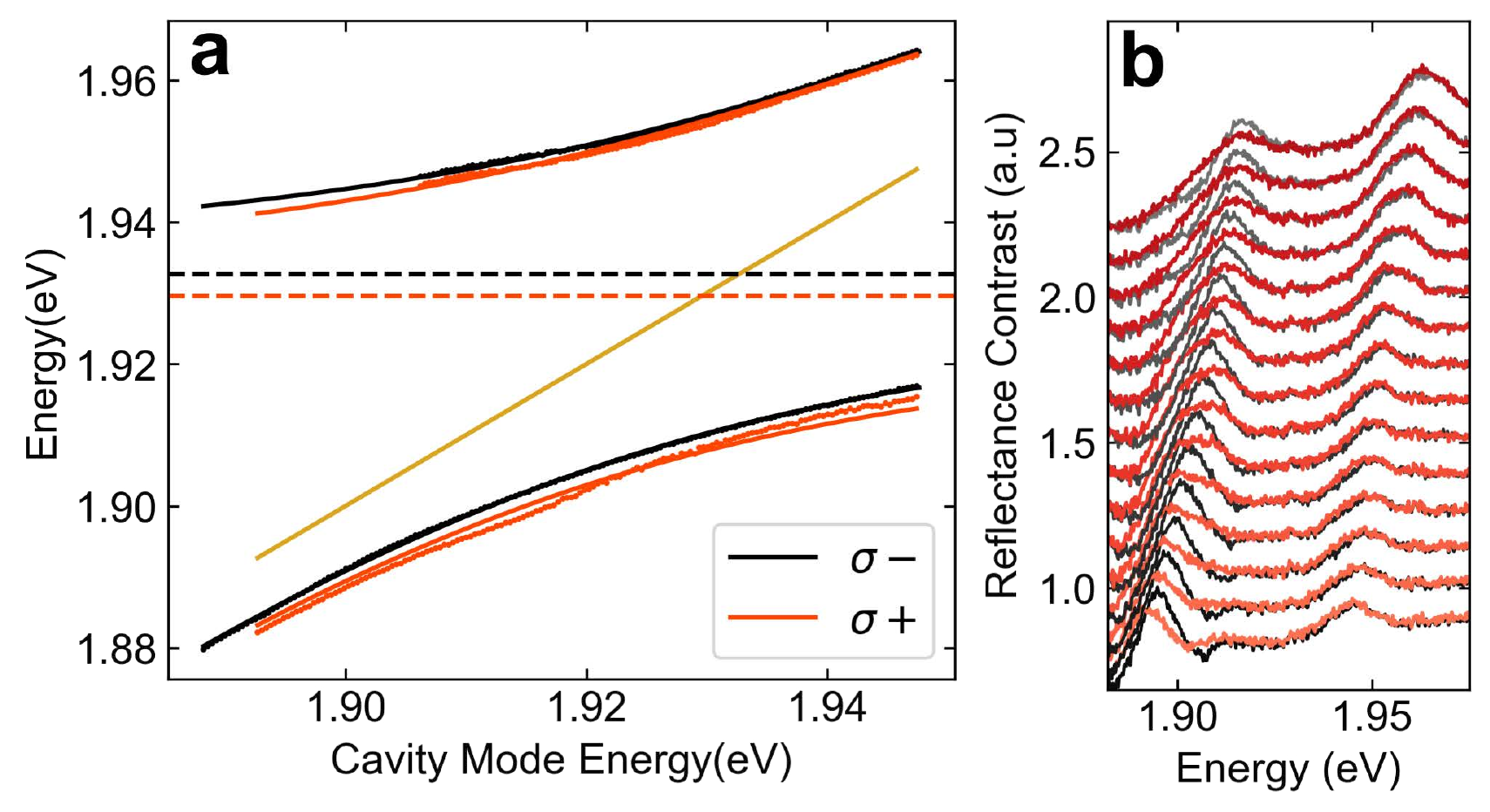}
\caption{\textbf{Zeeman Splitting of X\textsubscript{A}-polaritons. a)} Tunable cavity RC scans near X\textsubscript{A} energies.  Orange and black colours represent $\sigma^{+} $ and $\sigma^{-} $ detection respectively. Dots represent the extracted peak energies of individual spectra. The solid orange and black curves are the solutions to the fitted coupled oscillator Hamiltonians and the golden solid line is the extracted cavity mode energy. The extracted exciton energies are shown in the plot as the dashed horizontal lines of the corresponding colour. It can be seen that the Zeeman splitting of X\textsubscript{A} polaritons is of opposite sign to that of the hIX. The deviation of the $\sigma^{+}$ LPB datapoints from the coupled oscillator model solution (solid line) is due to the presence of the fully polarised trion at $\approx$ 1.91 eV. This can be seen as a broadening due to weak coupling of the lower energy peak of the orange ($\sigma^{+} $) spectra in \textbf{b)}, where a cascade plot of the spectra near resonance in the two polarizations is shown.   }\label{CO_DX_8T}
\end{figure}

In the presence of an out of plane magnetic field of 8~T, the same scans were repeated with unpolarised excitation and detecting  opposite circularly polarized light $\sigma^{+}$/$\sigma^{-}$ at each piezo voltage. Fig. \ref{CO_DX_8T} shows the coupled oscillator fits with data from $\sigma^{+}$ and $\sigma^{-}$ detection of the X\textsubscript{A} scan. It can be seen that the X\textsubscript{A}-polaritons exhibit an opposite sign of Zeeman splitting compared to hIX-polaritons (shown in the main text). 

\begin{figure}[pt]
\centering
\includegraphics[width=0.5\textwidth]{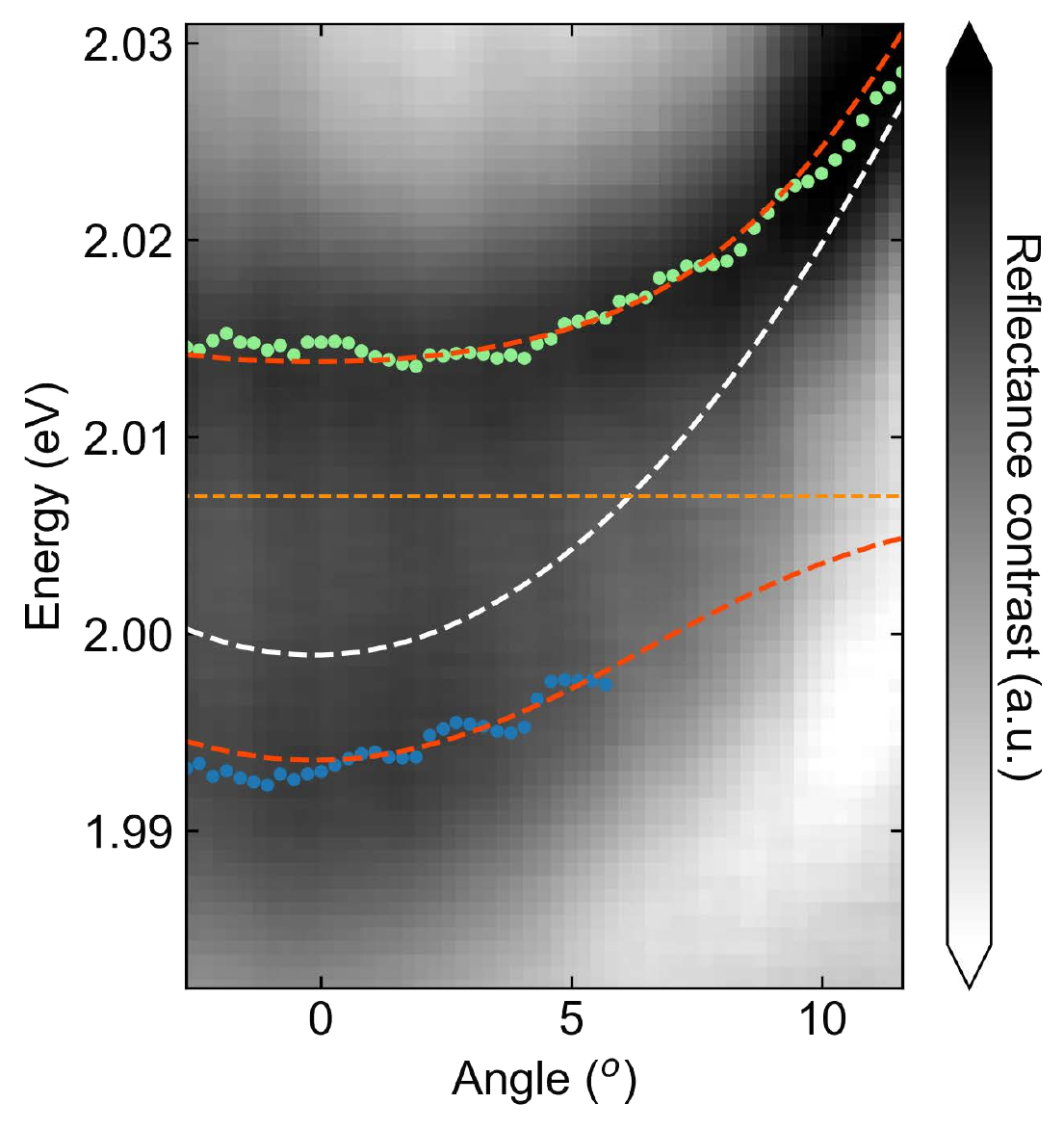}
\caption{\textbf{Coupled oscillator model fits on monolithic cavity.} Monolithic cavity dispersion near hIX with couple oscillator model fits. Blue and green points show the extracted peak positions of the LPB and UPB, respectively, from the spectra at each angle. These are then fitted to the two level coupled oscillator model and the solutions LPB and UPB are shown as dark orange dashed curves. The extracted cavity mode and exciton energy are shown as white and light orange dashed curves, respectively. }\label{CO_mono}
\end{figure}

Coupled oscillator models were also used for our monolithic cavity sample whose angular dispersion was measured with Fourier space imaging, as mentioned in the main text. The colour map with the extracted data points is presented in Fig. \ref{CO_mono}, with the overlaid couple oscillator fits.

\newpage

\section{Supplementary Note S3: Pump-probe resonant spectroscopy}\label{PP} 

\begin{figure}[t]
\centering
\includegraphics[width=1\textwidth]{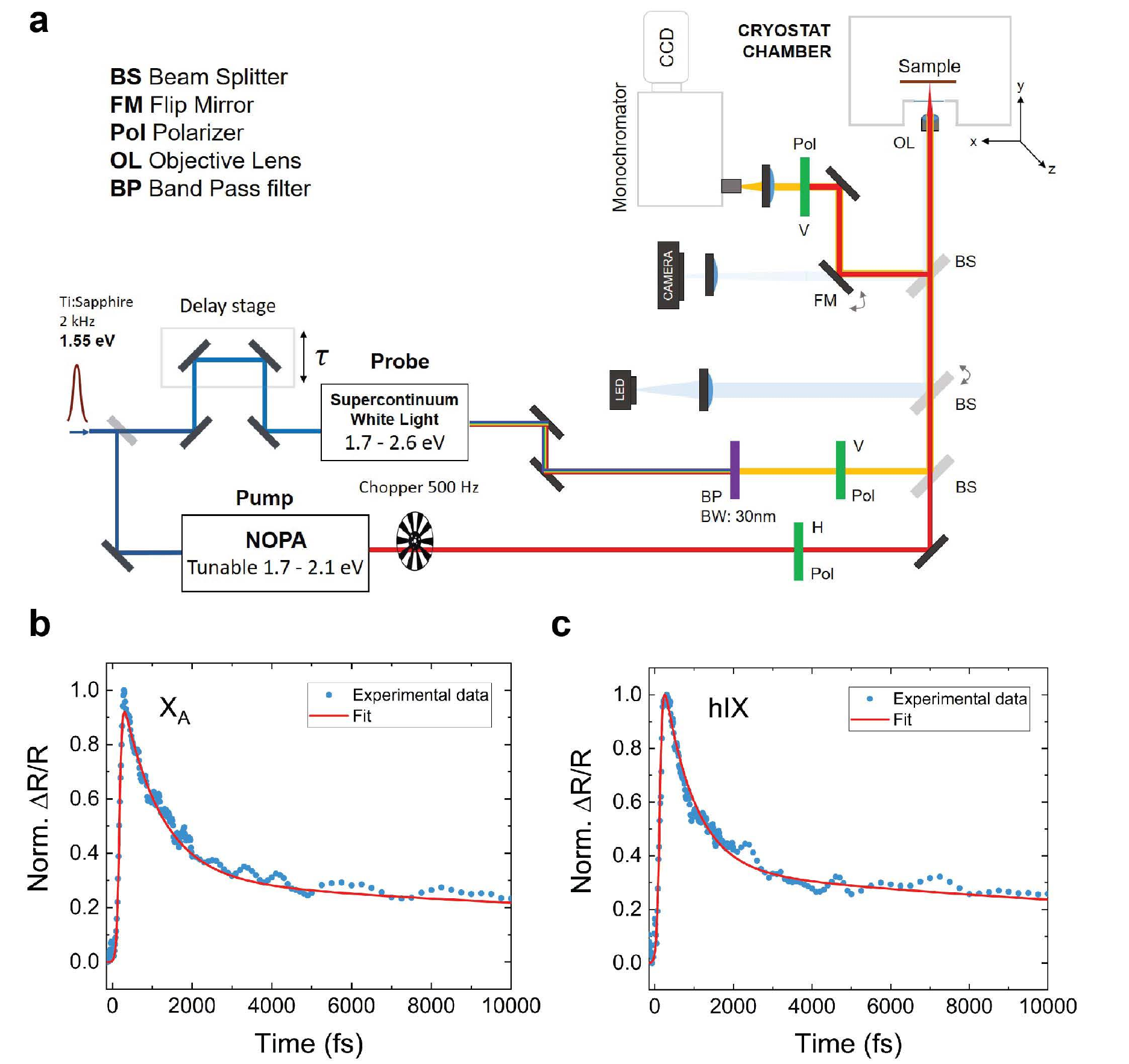}
\caption{\textbf{a)} Schematics of the pump-probe microscopy setup used for the experiments on the MoS\textsubscript{2} bilayers. \textbf{b, c)} Transient reflectivity traces for X\textsubscript{A} (b) and hIX (c) taken at 643 nm and 622 nm respectively. The red curves refer to the fitted bi-exponential decay function.}\label{figSI_PP}
\end{figure}

We performed time-resolved resonant pump-probe spectroscopy on the encapsulated BL MoS\textsubscript{2} out of the cavity, to measure the X\textsubscript{A} and hIX lifetimes in our system. Due to the small size of the samples, a microscopy setup allowing transient reflection measurements at low temperatures has been employed (Fig. \ref{figSI_PP} (a)). The setup is powered by an amplified Ti:sapphire laser (Coherent Libra) generating 100 fs pulses at 800 nm (1.55~eV) with 2 mJ pulse energy and 2 kHz repetition rate. A fraction of the laser output is used to seed a non-collinear optical parametric amplifier (NOPA) in the visible energy range. The generated pump pulses are modulated by a mechanical chopper at 500 Hz frequency. The broad-band probe pulses consist of a white-light continuum (WLC), generated from a 1-mm thick sapphire plate pumped by focusing the 800 nm output of the main laser. Pump and probe pulses are synchronized by means of a motorized delay stage. Pump and probe beams are then collinearly combined by a thin dichroic beam splitter and focused on the sample using an objective lens (NA=0.3), resulting in a 4$\mu m$ spot size. The samples are placed in a closed-cycle helium cryostat reaching a temperature of 6 K. The spatial overlap of the sample with the pump and probe spots is obtained by a three-axis (xyz) mechanical translation stage coupled to a home-built imaging system. After the interaction with the excited sample, the reflected probe pulse is collimated by the same objective lens and then sent to a spectrometer equipped with an electronically cooled Si CCD, to measure the differential reflection ($\Delta R/R$) signal.  

Figure \ref{figSI_PP} (b,c) shows the temporal exciton dynamics measured as the absorption bleaching signal in transient reflectivity, taken at the peak wavelengths of the pump-probe traces, 643 nm and 622 nm for X\textsubscript{A} and hIX respectively. For this experiment narrow-band pump pulses (FWHM=10nm) are tuned in resonance with each probed exciton and cross-polarized with respect to the probe pulses. The pump beam is then filtered by an additional polariser placed in the detection path before reaching the CCD. The probe spectral window has a narrow bandwidth of 30~nm with a central wavelength fixed at each exciton peak wavelength. The excitons decay traces show a fast component more prominent than the slow one, and can be fitted with a bi-exponential function convoluted with a Gaussian, taking into account the instrument response function. The resulting decay times are $\tau \textsubscript{fast}\approx950$~fs, $\tau \textsubscript{slow}\approx20$~ps for X\textsubscript{A}, and $\tau \textsubscript{fast}\approx780$~fs, $\tau \textsubscript{slow}\approx18$~ps for hIX. Considering the degeneracy of pump and probe energies in our experiments, the fast decay can be attributed to electron-phonon scattering processes from the K point to the lowest energy Q point of the Brillouin zone \cite{din2021ultrafast, nie2014ultrafast}, while the slow component is possibly related to radiative \cite{palummo2015exciton} or defect-mediated non-radiative recombination \cite{wang2015ultrafast}. We conclude that in our MoS\textsubscript{2} bilayer sample the fast and slow decay times for both X\textsubscript{A} and hIX are much longer than the temporal width of the probe pulses ($\approx 150$ fs).
\newpage

\section{Supplementary Note S4: Density estimation}\label{DC}

Exciton and polariton densities were calculated using an experimental approach considering a convolution of the laser profile and the Reflectance Contrast spectra, as in \cite{zhang2021van}.  Reflectance contrast, A$_{res}$, represents with good approximation the absorption of the each excitonic/polaritonic resonance \cite{zhang2021van}. Power absorbed by each exciton/polariton, ${P}_{{\rm{res}}}$, can be calculated as 
\begin{equation}
\label{eq:Pr_es}
{P}_{{\rm{res}}}=\frac{P\int L(E){{\rm{A}}}_{{\rm{res}}}(E){\rm{d}}E}{{I}_{{\rm{l}}{\rm{a}}{\rm{s}}{\rm{e}}{\rm{r}}}},
\end{equation}
where $P$ is the experimentally measured power, $\int L(E){{\rm{A}}}_{{\rm{res}}}(E){\rm{d}}E$ is the convolution of the laser spectrum profile, L(E), and the Reflectance Contrast spectrum in the range of energies of the resonant transition peak and $I_{\rm{laser}}=\int L(E){\rm{d}}E$ is the laser spectrum integrated intensity.  

The expression \eqref{eq:Pr_es} can then be used to calculate the particle density, $n_{\rm{res}}$, considering the laser repetition rate, $R_{\rm{laser}}$, laser spot size, $S_{\rm{laser}}$, and the excitation central energy, $E_{\rm{res}}$. Explicitly, an estimate for the density reads
\begin{equation}
    n_{\rm{res}}=\frac{{P}_{{\rm{res}}}}{R_{\rm{laser}}E_{\rm{res}}S_{\rm{laser}}}.
\end{equation}
We note that for the polariton density estimations, since A$_{res}$ is dependent on the angle, ${P}_{{\rm{res}}}$ was calculated integrating all the spectral quantities in both the energy and angular range of LPB and UPB. As such, the densities calculated with this procedure are the total polariton densities, which consider both LPB and UPB. 

For the values of polariton/exciton densities the error, $\rm{\epsilon_{n}} = \sqrt{{\left(\frac{{\rm{\epsilon }}_{{P}_{\rm{res}}}}{{P}_{{\rm{res}}}}\right)}^{2}+{\left(\frac{{\rm{\epsilon}}_{{E}_{{\rm{res}}}}}{{E}_{{\rm{res}}}}\right)}^{2}}$, is propagated with respect to standard error analysis rules \cite{hughes2010measurements}

\section{Supplementary Note S5: Comparison of hIX and X\textsubscript{A} nonlinearity under NB illumination}

\begin{figure}[h]
    \centering
    \includegraphics[width=0.4\textwidth]{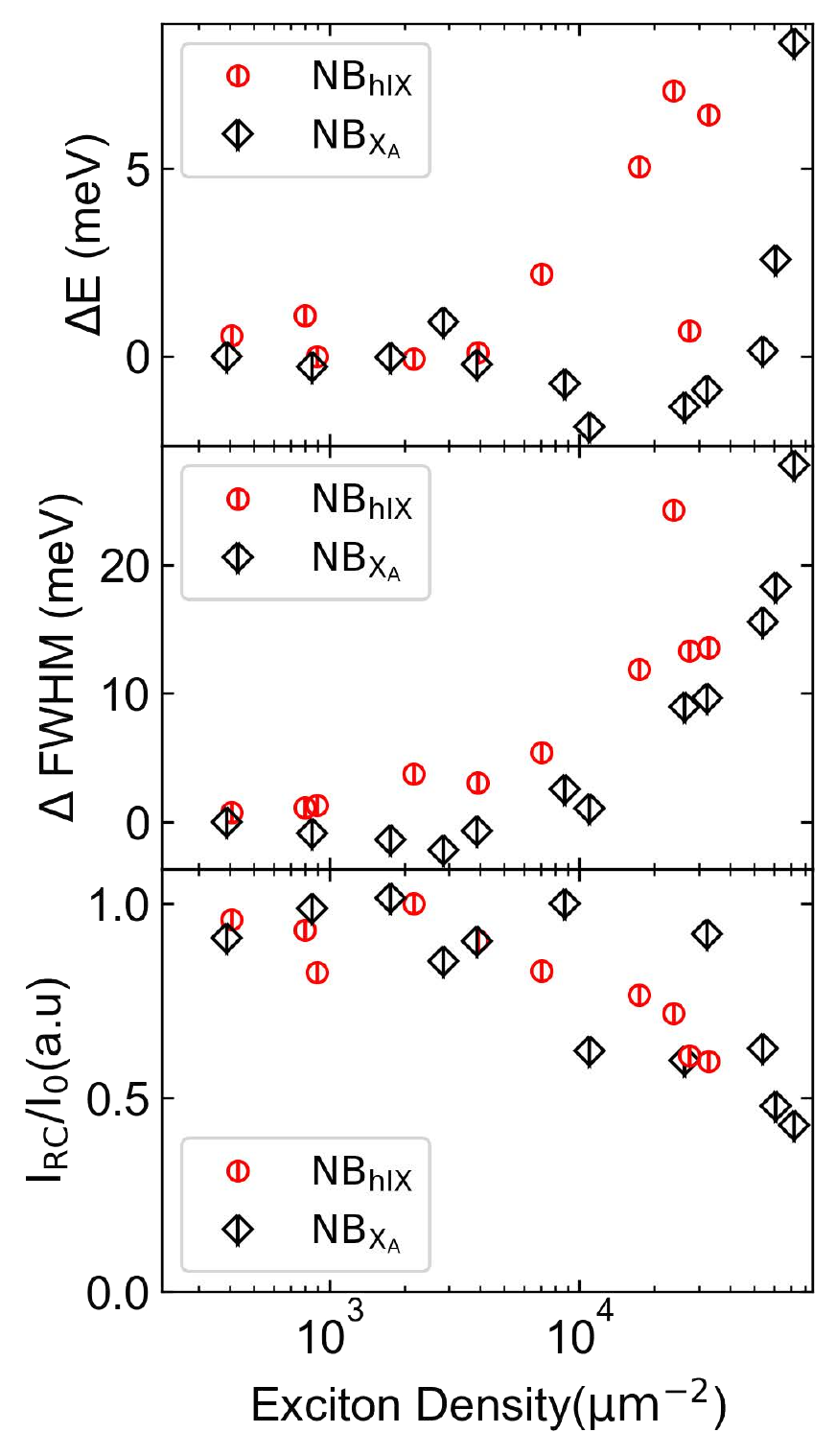}
    \caption{From top to bottom: plots of the energy shift ($\Delta E$), linewidth variation ($\Delta$FWHM) and normalised integrated intensity ($I_{\mathrm{RC}}/I_0$, where$I_0$ is the maximum integrated intensity) as a function density for  hIX (red marks in Fig. 3\textbf{d}) and X\textsubscript{A} (black marks in Fig. 3\textbf{e}) peaks measured in RC. hIX is bleaching, broadening and blueshifting at slightly lower densities compared to X\textsubscript{A}}
    \label{XAvshIX_NB}
\end{figure}

As mentioned in the text, the nonlinear behaviour of hIX is a slightly enhanced compared to X\textsubscript{A}. The Supplementary Figure \ref{XAvshIX_NB} shows the data of Fig. 3 in the main text such that a direct comparison between the two exciton species under separate narrow band (NB) illumination can be evaluated.

\section{Supplementary Note S6: hIX  and X\textsubscript{A} spectral fluence dependencies}\label{SF} 

Data shown in Fig. 3 of the main text are presented here in Supplementary Figure \ref{SI_SF} as a function of the raw spectral fluence. We define spectral fluence as the experimentally measured fluence normalised by the illumination spectral width in electron volts.

\begin{figure}[!h]
\centering
\includegraphics[width=0.7\textwidth]{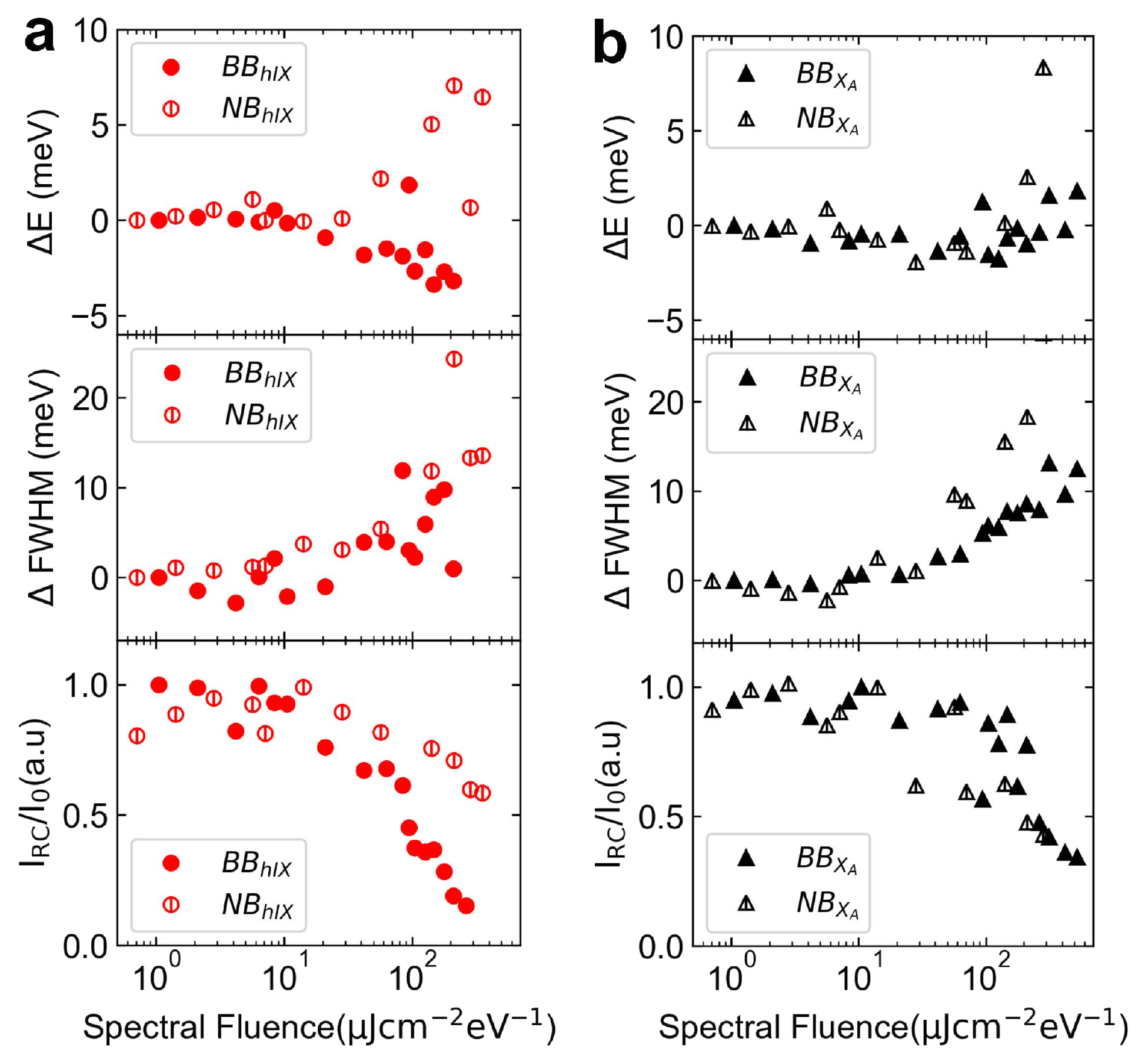}
\caption{\textbf{MoS\textsubscript{2} bilayer excitons nonlinear behaviour against spectral fluence} {\bf a), b)} From top to bottom: plots of the energy shift ($\Delta E$), linewidth variation ($\Delta$FWHM) and normalised integrated intensity ($I_{\mathrm{RC}}/I_0$, where $I_0$ is the maximum integrated intensity) as a function of the incident fluence normalised by the spectral width (Spectral Fluence) for hIX (red marks in \textbf{d}) and X\textsubscript{A} (black marks in \textbf{e}) peaks measured in RC. Solid marks relate to the experiments with BB excitation, covering both the hIX and the X\textsubscript{A}, while open marks refer to the NB excitation, at either the hIX or the X\textsubscript{A} energy. }\label{SI_SF}
\end{figure}

\section{Supplementary Note S7: Monolayer MoS\textsubscript{2} excitons nonlinear behaviour}\label{ML}

The density dependent nonlinearity was studied for an encapsulated MoS\textsubscript{2} monolayer on an identical DBR substrate outside the cavity, in narrow band (NB) illumination regime ($\approx$20 nm bandwidth). The results are then compared against the bilayer excitons excited with NB illumination. A less significant bleaching of excitons in the monolayer is clearly apparent, confirming the theoretical predictions of increased interactions in bilayer excitons. Extrapolation such data, we estimate that the monolayer to reach complete bleaching at up to an order of magnitude higher densities.  
\begin{figure}[h]
\centering
\includegraphics[width=0.92\textwidth]{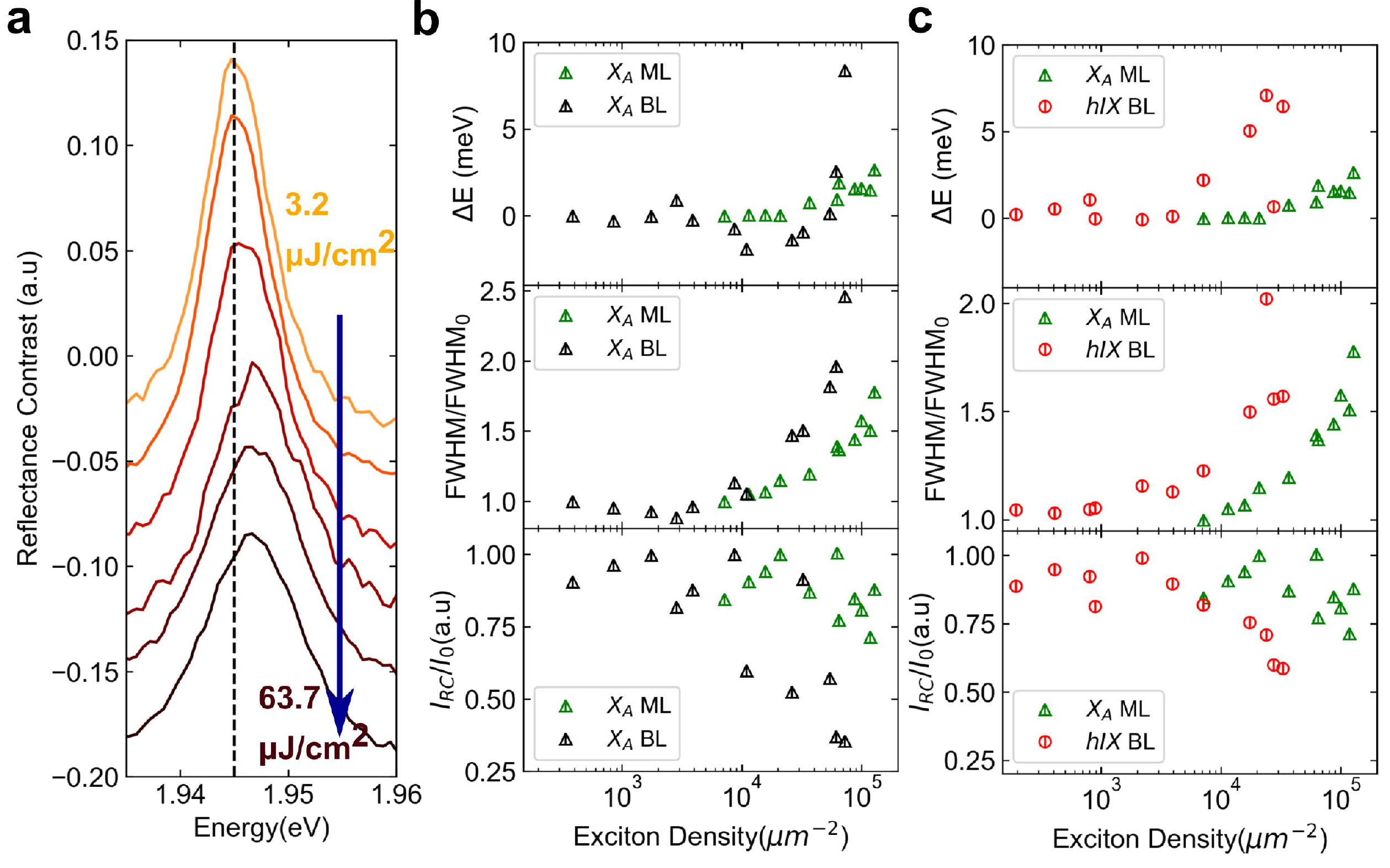}
\caption{\textbf{MoS\textsubscript{2} monolayer excitons nonlinear behaviour. a)} Waterfall of monolayer X\textsubscript{A} RC spectra with the darker colours representing larger pump powers. Despite a blueshift is apparent,, the exciton bleaching is much less pronounced than for the bilayer excitons. \textbf{b, c)} Nonlinearity comparison of monolayer X\textsubscript{A} to bilayer X\textsubscript{A} \textbf{(b)} and hIX \textbf{(c)}. From top to bottom: plots of the energy shift ($\Delta E$), normalised ($FWHM/FWHM
_{0}$) and normalised integrated intensity ($I_{\mathrm{RC}}/I_0$, where$I_0$ is the maximum integrated intensity) as a function of exciton density, for the monolayer X\textsubscript{A} (green), bilayer X\textsubscript{A}(black) and hIX (red) under narrow band excitation.}\label{SI_ML}
\end{figure}

\section{Supplementary Note S8: Theoretical discussion on interaction constants}
In our work we observed several nonlinear effects with contributions that depend on excitation conditions. In the BB regime, we already discussed the presence of two main species (flavours) of particles, namely direct X and indirect hIX excitons, determining the features of the sample's optical response. This can be seen from the reflectance spectra in {\color{blue} Fig. 2(a) and (b) of main text}. We observe both the energy shifts (discussed above), and additional bleaching of the peaks. This hints that the presence of conservative nonlinear processes (energy shifts) is accompanied by dissipative nonlinear processes. Below, we discuss various contributions, including the Coulomb-mediated scattering, optical saturation due to phase space filling, and nonlinear change of non-radiative decay and dephasing processes.

We stress that in general all the aforementioned processes contribute to the spectral signal we observed. The peak shape is due to the competing contribution of radiative and non-radiative decay processes. Both radiative $\Gamma_{\mathrm{R}}$ and non-radiative $\Gamma_{\mathrm{NR}}$ rates depend on the particle densities $n_{\mathrm D}$ and $n_{\mathrm I}$, for direct and interlayer excitons respectively. Here we simply refer to some generic density $n$, without specifying the particle flavour involved, as similar consideration apply to both. We relate the exciton radiative decay rate and the Rabi frequency $\Omega$ for polaritons, as they are both proportional to the particle oscillator strength $\Gamma_{\mathrm{R}}, \Omega \propto f_{\mathrm{osc}}$ \cite{Deng2010}. More accurately, we know that $\Gamma_{\mathrm{R}} \propto f_{\mathrm{osc}}$, and $\Omega(n) \propto \sqrt{f_{\mathrm{osc}}}$. With  $\Omega(n) = \Omega_0 \sqrt{1 - \xi_{\mathrm{sat}} n}$, we write the relation: $\Gamma_{\mathrm R}(n) = \Gamma_{\mathrm 0} - g_{\mathrm{sat}} n$, with $g_{\mathrm{sat}}$ being a saturation constant to be determined ($g_{\mathrm{sat}} \propto \xi_{\mathrm{sat}}$). As consequence of geometrical properties of excitons, the saturation factor shape $\xi_{\mathrm{sat}} \simeq 4 \alpha^2$ \citep{Emmanuele2020}, where $\alpha$ is the exciton Bohr radius.

In parallel to this process, non-radiative processes play a major role in bleaching. As there are many decay and dephasing channels, the full treatment of possible processes if formidable. Here, we address as the key effect the decay process due to Coulomb scattering \cite{Erkensten2021}. As a result, the Coulomb scattering induced decay is proportional to the Coulomb scattering matrix
\begin{equation}
    \Gamma_{\mathrm{NR}} \propto \sum_{\mathbf q \neq 0} \vert V^{\mathrm{dir}}(\mathbf{q}) - V^{\mathrm{exch}}(\mathbf{q}) \vert^2 \delta(E(\mathbf{q}) + E(-\mathbf{q}) - 2 E(\mathbf Q \sim 0)),
\end{equation}
where $E$ is the energy of involved particles and $V^{\mathrm{dir, exch}}(\mathbf{q})$ are direct and exchange particle scattering amplitudes, as discussed in Sec. S9. Note that in the main text for brevity we refer to the combined effect of different Coulomb-based processes using the combined interaction constant $V_\mathrm{Coul}$. 
The discussed decay properties of generic particles combine into the final shape of spectral peak $\mathcal L(E)$, {\color{blue} see Fig. 4(a)} \cite{Ivchenko1996}:
\begin{equation}
 \label{eq:lorenzian}
    \mathcal L(E) = \frac{1}{\pi} \frac{\Gamma_{\mathrm R}(n)^2}{ (E - E_0)^2 + [\Gamma_{\mathrm R}(n) + \Gamma_{\mathrm{NR}}(n)]^2},
\end{equation}
where $E_0$ is the position of the peak, $\Gamma_{\mathrm R}$ is the radiative decay rate and $\Gamma_{\mathrm{NR}}$ is the non-radiative rate. Analysing the experimental data with minimal square method, we estimate the non-radiative decay rate to be of the order of $\sim 1$ meV for direct excitons and $\sim 10$ meV for indirect excitons. The ratio between the two agrees well with the estimates for the interaction constants (see the discussion below). With combining both radiative and non-radiative bleaching, we find the experimental data are well described by the relation
\begin{equation}
    \Omega(n) = \sqrt{\Omega_0^2 (1 - \xi_{\mathrm{sat}} \alpha^2 \; n)^2 - \Gamma_{\mathrm{NR}}^2 (1 + \xi_{\mathrm{NR}} \alpha^2 \; n)^2},
\end{equation}
where the discussed experimental observation are well described by $\xi_{\mathrm{NR}} \sim 10$, and $\xi_{\mathrm{sat}} \sim 7$ describes the nonlinear saturation of the Rabi splitting. The origin of the saturation term comes from the interlayer exciton phase space filling, and is reminiscent to phase space filling effects discussed in Ref. \cite{Emmanuele2020}.

Finally, let us consider the exciton-exciton Coulomb scattering. We provide here the estimates for interaction constants, with considering our gained knowledge of the effective interlayer distance $d$ and particles Bohr radii $\alpha_{\mathrm{D,I}}$. We build the interaction by following the procedure in Refs.~\cite{Kyriienko2012, Ciuti1998}. As a signature of particle indistinguishability, we have to consider both direct scattering $V^{\mathrm{dir}}(\mathbf q)$ and exchange scattering $V^{\mathrm{exch}}(\mathbf q)$ amplitudes, with $\mathbf q$ being the exchanged momentum. We omitted particle momenta as we consider total momentum to be zero. Explicitly, the two contributions read
\begin{equation} 
\begin{split}
    V^{\mathrm{dir}}(\mathbf q ) =& \int d^2 r_{\mathrm e} \, d^2 r_{\mathrm h} \, d^2 r_{\mathrm e^\prime} \, d^2 r_{\mathrm e^\prime}
    \phi^\ast(\mathbf r_{\mathrm e}, \mathbf r_{\mathrm h})\phi^\ast(\mathbf r_{\mathrm e^\prime}, \mathbf r_{\mathrm h^\prime})
    \\
    &
    \Phi^{\mathrm{tot}}(\mathbf r_{\mathrm e}, \mathbf r_{\mathrm h}, \mathbf r_{e^\prime}, \mathbf r_{\mathrm h^\prime}) \phi(\mathbf r_{\mathrm e}, \mathbf r_{\mathrm h}) \phi(\mathbf r_{e^\prime}, \mathbf r_{\mathrm h^\prime}),
\end{split}
\end{equation}
and
\begin{equation} 
\begin{split}
    V^{\mathrm{exch}}(\mathbf q ) =& \int d^2 r_{\mathrm e} \, d^2 r_{\mathrm h} \, d^2 r_{\mathrm e^\prime} \, d^2 r_{\mathrm e^\prime}
    \phi^\ast(\mathbf r_{\mathrm e}, \mathbf r_{\mathrm h})\phi^\ast(\mathbf r_{\mathrm e^\prime}, \mathbf r_{\mathrm h^\prime})
    \\
    &
    \Phi^{\mathrm{tot}}(\mathbf r_{\mathrm e}, \mathbf r_{\mathrm h}, \mathbf r_{e^\prime}, \mathbf r_{\mathrm h^\prime}) \phi(\mathbf r_{e^\prime}, \mathbf r_{\mathrm h}) \phi(\mathbf r_{\mathrm e}, \mathbf r_{\mathrm h^\prime}),
\end{split}
\end{equation}
where we define $\Phi^{\mathrm{tot}}(\mathbf r_{\mathrm e}, \mathbf r_{\mathrm h}, \mathbf r_{e^\prime}, \mathbf r_{\mathrm h^\prime})$ as the sum of mutual particle interaction. For the scattering elements above, we shall consider two separate cases for direct and indirect excitons, as both wavefunctions and Coulomb terms differ. 
\begin{figure}[ht!]
    \centering
    \includegraphics[width=0.9\linewidth]{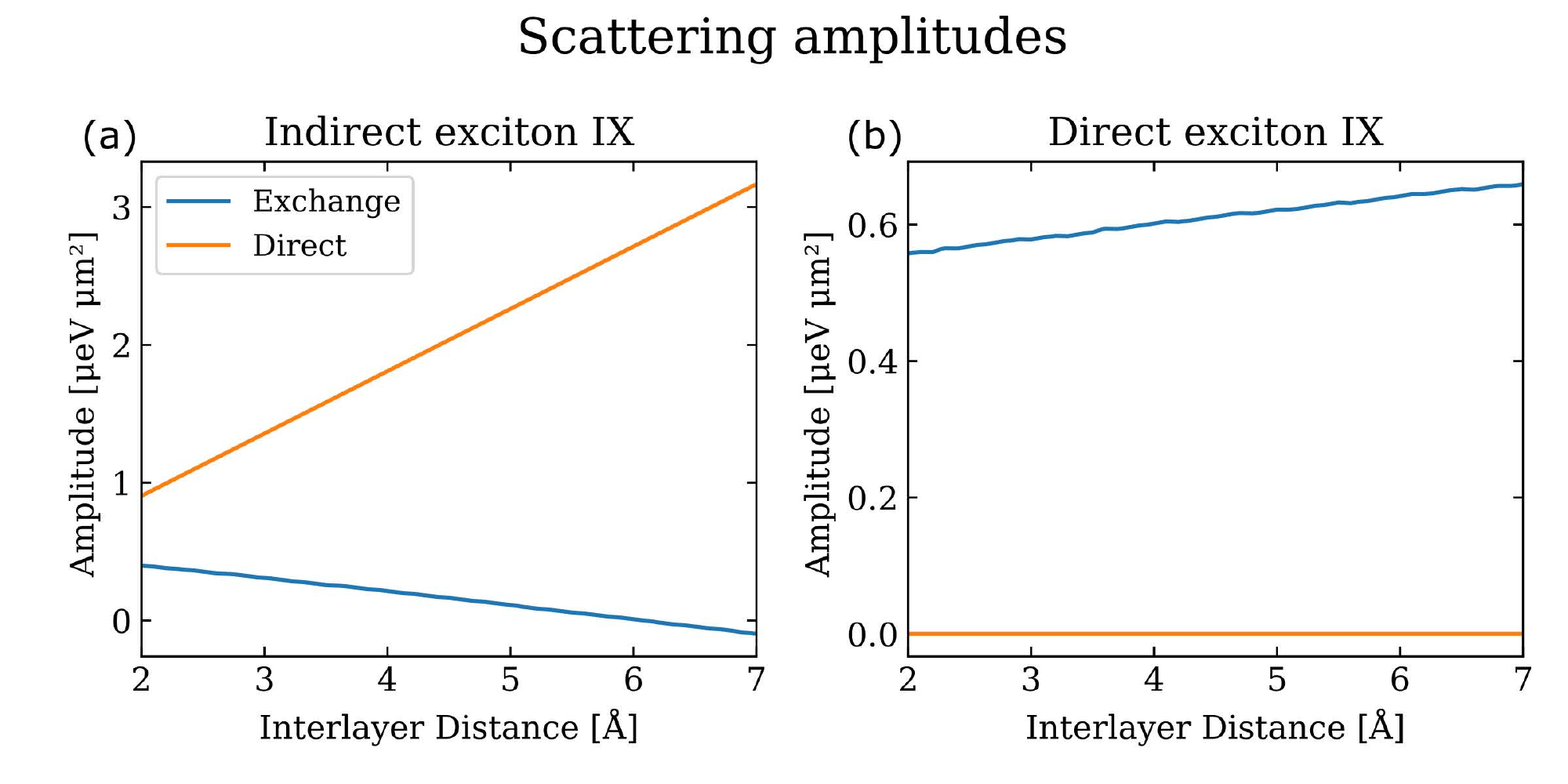}
    \caption{\textbf{Scattering amplitudes for exciton-exciton Coulomb interaction. (a,b)} We show the change of scattering amplitudes with the interlayer distance. In \textbf{(a)}, note the behavior of exchange scattering amplitude for indirect excitons, becoming negative past a threshold distance. Direct scattering amplitude grows linearly with distance. \textbf{(b)} We plot characteristic scattering amplitudes for direct excitons, where the direct process is zero at the negligible exchanged momentum (orange line), and the blue line corresponds to the exchange processes.}
    \label{fig:nonLinearities}
\end{figure}

With direct excitons, we have $V^{\mathrm{dir}}_{\mathrm {X-X}}(\mathbf q = \mathbf 0 ) = 0$, and
\begin{equation}
\begin{split}
    V^{\mathrm{exch}}_{\mathrm {X-X}}(\mathbf q = \mathbf 0 ) =& \left({2 \over \pi}\right)^2 {e^2 \over 4 \pi \epsilon \epsilon_0} \alpha_{D} \int dx \, dy \, d \theta {2 \pi x y \over \delta(x,y, \theta)}
    \\
    &
    {1 \over (1 + r_{\mathrm 0}\delta(x, y, \theta) / \alpha_{\mathrm D} )^2 - (\delta(x, y,  \theta) r_{\mathrm{0}}/\alpha_{\mathrm D})^2 e^{- 2 \delta(x,y,\theta){d \over \alpha_{\mathrm D}}}} \cdot
    \\
    &
     \left( 1 + {r_{\mathrm 0} \over \alpha_{\mathrm D} }\left(1 - e^{- 2 \delta(x,y, \theta) {d \over \alpha_{\mathrm D}}} \right) \right)\cdot \biggl( {(-1) \over (1 + x^2)^3}  {1 \over (1 + y^2)^3}
    \\
    &
    + {1 \over (1 + x^2)^3} {e^{- \delta(x,y,\theta) {d \over \alpha_{\mathrm D}}} \over (1 + y^2)^{3/2} (1 + x^2)^{3/2}} \biggr),
    \end{split}
\end{equation}
where $\delta(x, y, \theta) = \sqrt{x^2 + y^2 - 2 x y \cos{\theta}}$. With indirect excitons, the direct scattering amplitude recovers the capacitor formula $V^{\mathrm{dir}}_{\mathrm {I-I}}(\mathbf q = \mathbf 0 ) = {e^2 / (\epsilon \epsilon_0 A)} d$, where $A$ is the sample area. Finally, the indirect exciton exchange potential has to be evaluated as
\begin{equation}
\begin{split}
    V^{\mathrm{exch}}_{\mathrm {I-I}}(\mathbf q = \mathbf 0 ) =& \left({2 \over \pi}\right)^2 {e^2 \over 4 \pi \epsilon \epsilon_0} \alpha_{D} \int dx \, dy \, d \theta {2 \pi x y \over \delta(x,y, \theta)}
    \\
    &
    {1 \over (1 + r_{\mathrm 0}\delta(x, y, \theta) / \alpha_{\mathrm I} )^2 - (\delta(x, y,  \theta) r_{\mathrm{0}}/\alpha_{\mathrm I})^2 e^{- 2 \delta(x,y,\theta){d \over \alpha_{\mathrm I}}}} \cdot 
    \\
    &
    \biggl[ {(-1) \over (1 + x^2)^3}  {1 \over (1 + y^2)^3} \left( 1 + {r_{\mathrm 0} \over \alpha_{\mathrm I} }\left(1 - e^{- 2 \delta(x,y, \theta) {d \over \alpha_{\mathrm I}}} \right) \right)
    \\
    &
    + {1 \over (1 + x^2)^3} {e^{- \delta(x,y,\theta) {d \over \alpha_{\mathrm I}}} \over (1 + y^2)^{3/2} (1 + x^2)^{3/2}} \biggr].
    \end{split}
\end{equation}
Figs.~\ref{fig:nonLinearities}(a) and (b) show the dependence of scattering amplitudes on the interlayer distance. We note the behavior of exchange scattering amplitude for indirect excitons, becoming negative past threshold distance (see Fig.~\ref{fig:nonLinearities}(a)). Characteristically, for intralayer (i.e. direct) excitons we have a zero direct scattering amplitude, with non-zero contributions only due to the particle exchange process.
With the estimated parameters, we find interlayer exciton-exciton interaction to have a scattering constant of $V_{\mathrm{I-I}}(\mathbf q = \mathbf 0 ) \simeq 2.5$~$\mu$eV~$\mu$m$^2$, setting the scale for the Coulomb-based interactions. 
%

\section{Supplementary Note S9: Theory for energy shift}\label{sec:energyShift}
The out-of-cavity results {\color{blue} (see Fig. 3 main text)} reveal an opposite behavior of the sample response depending on the excitation regimes, where narrow bandwidth (NB) and the broad bandwidth (BB) regimes are considered. In the NB case, we observe a blueshift for the hIX peak, as already reported in literature for dipolar excitons \cite{Schindler2008, Zimmermann2007, Kyriienko2012}, while in the BB regime we observe a redshifted signal. The latter is unexpected, as in the system with dipolar excitons and predominantly positive scattering matrix elements for exchange terms, the emergence of some effective attractive nonlinearities requires a new possible mechanism. Below, we motivate the emergence of such a mechanism unique to the bilayer system. 

First, let us analyse the energy of system as the expectation value of the full Hamiltonian $\hat{\mathcal{H}}$. We consider electrons and holes distributed over bilayer as shown in {\color{blue}Fig. 1(d) [main text]}. We call $N_{\mathrm{I(D)}}$ the number of indirect (direct) excitons in the sample, and $n_{\mathrm{I(D)}}$ the particle density. The spin indices are omitted for brevity. The Hamiltonian of the system can be written as
\begin{equation}
    \hat{\mathcal H} = \hat{\mathcal H}_{\mathrm{T}} + \hat{\mathcal H}_{\mathrm{sm}} + \hat{\mathcal H}_{\mathrm{df}},
\end{equation}
where $\hat{\mathcal{H}}_{\mathrm{T}}$ is the kinetic term, $\hat{\mathcal H}_{\mathrm{sm}}$ and $\hat{\mathcal H}_{\mathrm{df}}$ are the Coulomb interactions. With $\hat{\mathcal H}_{\mathrm{sm}}$ we refer to interacting particles belonging to a same band, and $\hat{\mathcal H}_{\mathrm{df}}$ corresponds to different dispersion bands. Explicitly, the kinetic energy reads
\begin{equation}
    \hat{\mathcal H}_{\mathrm{T}} = \sum_{\mathbf k} \Big[ \varepsilon_{c}^t(\mathbf{k}) \hat a^\dagger_{ \mathbf{k}} \hat a_{ \mathbf{k}} + \varepsilon_{v}^t(\mathbf{k})b^\dagger_{\mathbf{k}}b_{\mathbf{k}} + \varepsilon_{c}^b(\mathbf{k})c^\dagger_{\mathbf{k}}c_{\mathbf{k}}\Big],
\end{equation}
where $\hat a^\dagger_{\mathbf k}$ and $\hat b^\dagger_{\mathbf k}$ are creation operators for conduction and valence bands of the top layer, respectively.  $\hat c^\dagger_{\mathbf k}$ is an electron annihilation operator for the conduction band of the bottom layer. Each operator is labelled with a crystal momentum $\mathbf k$. $\varepsilon_{\mathrm{c(v)}}^{\mathrm{t(b)}}(\mathbf k)$ are dispersions for conduction (valence) bands in the top (bottom) layer. The interactions are mediated through the Keldysh-Rytova potential. The corresponding interaction Hamiltonian reads
\begin{equation}
   \hat{\mathcal H}_{\mathrm{sm}} =     \frac{1}{2} \sum_{\mathbf{k}, \mathbf{k}^\prime, \mathbf{q}}\Big[ V^{\mathrm{intra}}_{\mathrm{KR}}(\mathbf{q}) ( a^\dagger_{\mathbf{k} - \mathbf{q}}\hat{a}^\dagger_{\mathbf{k}'+\mathbf{q}}\hat{a}_{\mathbf{k}'} \hat{a}_{\mathbf{k}} + \hat{b}^\dagger_{\mathbf{k} - \mathbf{q}}\hat{b}^\dagger_{\mathbf{k}'+\mathbf{q}}\hat{b}_{\mathbf{k}'} \hat{b}_{\mathbf{k}} + c^\dagger_{\mathbf{k} - \mathbf{q}}\hat{c}^\dagger_{\mathbf{k}'+\mathbf{q}}\hat{c}_{\mathbf{k}'} \hat{c}_{\mathbf{k}} )\Big],
\end{equation}
\begin{equation}
    \hat{\mathcal H}_{\mathrm{df}} = \sum_{\mathbf{k}, \mathbf{k}^\prime, \mathbf{q}}\Big[ V^{\mathrm{inter}}_{\mathrm{KR}}(\mathbf{q}) ( \hat{a}^\dagger_{\mathbf{k} - \mathbf{q}}\hat{c}^\dagger_{\mathbf{k}'+\mathbf{q}}\hat{c}_{\mathbf{k}'} \hat{a}_{\mathbf{k}} + \hat{b}^\dagger_{\mathbf{k} - \mathbf{q}}\hat{c}^\dagger_{\mathbf{k}'+\mathbf{q}}\hat{c}_{\mathbf{k}'} \hat{b}_{\mathbf{k}} ) + V^{\mathrm{intra}}_{\mathrm{KR}} \hat{a}^\dagger_{\mathbf{k} - \mathbf{q}}\hat{b}^\dagger_{\mathbf{k}'+\mathbf{q}}\hat{b}_{\mathbf{k}'} \hat{a}_{\mathbf{k}}\Big].
\end{equation}
We define the exciton creation operators in the form
\begin{align}
    \hat D_{\mathrm \mu}^\dagger(\mathbf Q) =& \sum_{\mathbf k} \phi^{\mathrm \mu}_{\mathrm D}(\mathbf k)  \hat{a}^\dagger_{\mathbf k + \gamma_{\mathrm e} \mathbf Q} \hat b_{\mathbf k - \gamma_{\mathrm h} \mathbf Q},
    \\
    \hat I_{\mathrm \nu}^\dagger(\mathbf P) =& \sum_{\mathbf k}\phi^{\mathrm \nu}_{\mathrm I}(\mathbf k) \hat{a}^\dagger_{\mathbf k + \gamma_{\mathrm e} \mathbf P} \hat c_{\mathbf k - \gamma_{\mathrm h} \mathbf P},
\end{align}
with $\hat D^\dagger_{\mathrm \mu}(\mathbf Q)$ [$\hat I^\dagger_{\mathrm \nu}(\mathbf P)$] and $\phi^{\mathrm \mu}_{\mathrm D}(\mathbf k)$ [$\phi^{\mathrm \nu}_{\mathrm I}(\mathbf k)$] being the direct [indirect] exciton creation operator and wave function, respectfully. $\mathbf Q, \mathbf P$ are crystal momenta, and $\mu, \nu$ are state indices. When we omit the $\mu$ and $\nu$ indices, and the total momentum, we refer the ground state at crystal momentum $\mathcal Q = 0$. To take into account for particle non-bosonicity and consequent nonlinear behaviour, we consider the expectation value of a system over a multi-particle state created by exciting the vacuum state $\vert \Omega_{\mathrm{0}} \rangle$. It includes $N_{\mathrm D}$ direct excitons and $N_{\mathrm I}$ indirect excitons. The expectation $ \langle \Omega_{\mathrm 0} \vert \hat D^{N_{\mathrm{D}}} \hat I^{N_{\mathrm{I}}} \hat{\mathcal H} \hat D^{\dagger N_{\mathrm{D}}} \hat I^{\dagger N_{\mathrm{I}}} \vert \Omega_{\mathrm{0}} \rangle$ then denotes the total energy of the many-body system. By following a procedure for accounting non-bosonic correction at increasing order~\cite{Combescot2008}, we first commute the Hamiltonian with the product of exciton operators $\hat D^{\dagger N_{\mathrm D}}$, leading to
\begin{equation}\label{eq:direct_comm}
\begin{split}
    [\mathcal H, \hat D^{\dagger N_{\mathrm D}}] =& N_{\mathrm{D}} E_{\mathrm{D}} \hat D^{\dagger N_\mathrm{D} - 1} + N_{\mathrm{D}} \hat D^{\dagger N_{\mathrm D} - 1} \hat V_{\mathrm D} 
    \\
    & + \frac{N_{\mathrm D}(N_{\mathrm D} - 1)}{2} \hat D^{\dagger N_{\mathrm D} - 2} \sum_{ \mu,\nu, \mathrm q} V^{\mu, \nu}_{\mathrm{D-D}}(\mathbf q) \hat D^\dagger_{\nu}( \mathbf q) \hat D^\dagger_{\mu}(- \mathbf q),
\end{split}
\end{equation}
where $V^{\mu, \nu}_{\mathrm{D-D}}(\mathbf q)$ is the scattering matrix element of two direct excitons exchanging a momentum of $\mathbf q$. $\hat V_{\mathrm D}$ is the scattering potential~\cite{Combescot2008}, which arises from the commutator $[\mathcal H, \hat D^\dagger]$ and is a signature of the phase space filling. Similarly, for indirect excitons we get
\begin{equation}{\label{eq:indirect_comm}}
\begin{split}
    [\mathcal H, \hat I^{\dagger N_{\mathrm I}}] =& N_{\mathrm{I}} E_{\mathrm{I}} \hat I^{\dagger \mathrm{N}_I - 1} + N_{\mathrm{I}} \hat I^{\dagger N_{\mathrm I} - 1} \hat V_{\mathrm I} 
    \\
    & + \frac{N_{\mathrm I}(N_{\mathrm I} - 1)}{2} \hat I^{\dagger \mathrm{N_{\mathrm I} - 2}} \sum_{ \mu,\nu, \mathrm q} V^{\mu, \nu}_{\mathrm{I-I}}(\mathbf q) \hat I^\dagger_{\nu}( \mathbf q) \hat I^\dagger_{\mu}(- \mathbf q),
\end{split}
\end{equation}
with the notation being similar to Eq.~\eqref{eq:direct_comm}. We note the property of the scattering potential such that $\hat V_{\mathrm{D, I}} \vert \Omega_0 \rangle = 0$. With the given commutators, we can rewrite the total energy as
\begin{equation}\label{eq:systemEnergy}
\begin{split}
    &\langle \Omega_{\mathrm{0}} \vert \hat D^{N_{\mathrm{D}}} \hat I^{N_{\mathrm{I}}} \hat{\mathcal H} \hat D^{\dagger N_{\mathrm{D}}} \hat I^{\dagger N_{\mathrm{I}}} \vert \Omega_{\mathrm{0}} \rangle =
    \\
    & = 
    N_{\mathrm{I}} E_{\mathrm{I}} \langle \Omega_{\mathrm{0}} \vert \hat D^{N_{\mathrm{D}}} \hat I^{N_{\mathrm{I}}} \hat D^{\dagger N_{\mathrm{D}}} \hat I^{\dagger N_{\mathrm{I}}} \vert \Omega_{\mathrm{0}} \rangle + 
    N_{\mathrm{D}} E_{\mathrm{D}} \langle \Omega_{\mathrm{0}} \vert \hat I^{N_{\mathrm{I}}} \hat D^{N_{\mathrm{D}}} \hat D^{\dagger N_{\mathrm{D}}} \hat I^{\dagger N_{\mathrm{I}}} \vert \Omega_{\mathrm{0}} \rangle
    \\
    & +
    \frac{N_{\mathrm I}(N_{\mathrm I} - 1)}{2} \hat I^{\dagger N_{\mathrm I} - 2} \sum_{ \mu,\nu, \mathrm q} V^{\mu, \nu}_{\mathrm{I-I}}(\mathbf q) \langle \Omega_{\mathrm{0}} \vert \hat D^{N_{\mathrm{D}}} \hat I^{N_{\mathrm{I}}} \hat D^{\dagger N_{\mathrm{D}}} \hat I^{\dagger N_{\mathrm{I}} - 2} \hat I^\dagger_{\nu}( \mathbf q) \hat I^\dagger_{\mu}(- \mathbf q) \vert \Omega_{\mathrm{0}} \rangle
    \\
    & +
    \frac{N_{\mathrm D}(N_{\mathrm D} - 1)}{2} \hat D^{\dagger N_{\mathrm D} - 2} \sum_{ \mu,\nu, \mathrm q} V^{\mu, \nu}_{\mathrm{D-D}}(\mathbf q) \langle \Omega_{\mathrm{0}} \vert \hat D^{N_{\mathrm{D}}} \hat I^{N_{\mathrm{I}}} \hat D^{\dagger N_{\mathrm{D}} - 2} \hat D^\dagger_{\nu}( \mathbf q) \hat D^\dagger_{\mu}(- \mathbf q) \hat I^{\dagger N_{\mathrm{I}}} \vert \Omega_{\mathrm{0}} \rangle
    \\
    & + 
    N_{\mathrm{D}} N_{\mathrm{I}} \sum_{\mu, \nu, \mathbf q} V^{\mu, \nu}_{\mathrm{D - I}}(\mathbf q) \langle \Omega_{\mathrm{0}} \vert \hat D^{N_{\mathrm{D}}} \hat I^{N_{\mathrm{I}}} \hat D^{\dagger N_{\mathrm{D}} - 1} \hat D^\dagger_{\mu}( \mathbf q) \hat I^\dagger_{\nu}(- \mathbf q) \hat I^{\dagger N_{\mathrm{I}} - 1} \vert \Omega_{\mathrm{0}} \rangle.
\end{split}
\end{equation}
Eq.~\eqref{eq:systemEnergy} describes various energy contributions (linear and nonlinear) that are present in the system. We note that each term is proportional to the expectation value $\langle \Omega_{\mathrm{0}} \vert \hat D^{N_{\mathrm{D}}} \hat I^{N_{\mathrm{I}}} \hat D^{\dagger N_{\mathrm{D}}} \hat I^{\dagger N_{\mathrm{I}}} \vert \Omega_{\mathrm{0}} \rangle$, which deviates from $1$ due to the non-bosonicity of composite excitons. This reveals three possible creation potentials $\hat V^{\mu, \nu}_{\mathrm{D-D}}, \hat V^{\mu, \nu}_{\mathrm{I-I}}$, and $\hat V^{\mu, \nu}_{\mathrm{D-I}}$, which noticeably generate cross-flavour interaction. The presence of creation potentials, as well as quadratic scaling of these terms in the total energy, is the signature of nonlinear behaviour. Terms in lines 2 and 3 of Eq.~\eqref{eq:systemEnergy} are the energy shifts for direct and indirect excitons due to phase space filling within the same exciton flavour. The last term in line 4 is the cross interaction of direct and indirect excitons, allowing for extra energy shifts in modes that are not statistically independent. Namely, the corresponding operators for intralayer and interlayer excitons do not commute as they share a hole, but formed by different electrons in conduction bands. This leads to the negative valued commutator, and here we identify the origin of unusual redshift, as observed in the work. We refer to this effect as hole crowding, where hole population being shared between the two different particles (hIX and X\textsubscript{A}). This is different from statistical deviation of same excitons, where both carriers are exchanged, which we simply refer as interlayer exciton phase space filling, in analogy to the monolayer case. The effect of intra-flavour terms has already been discussed in Ref.~\cite{Emmanuele2020}, and leads to positive energy shifts. However, the cross-flavour terms only emerge in bilayers, and to date remained unexplored. We evaluate the considered term in $\mathbf q \simeq 0$, as it gives the dominant contribution observed in experiments, and we consider particles being in the ground state as a main occupation at low temperatures. With this, we rewrite the mutual energy shift $\Delta E_{\mathrm I} = \sum_{\mu, \nu, \mathbf q} V^{\mu, \nu}_{\mathrm{D - I}}(\mathbf q) \langle \Omega_{\mathrm{0}} \vert \hat D^{N_{\mathrm{D}}} \hat I^{N_{\mathrm{I}}} \hat D^{\dagger N_{\mathrm{D}} - 1} \hat D^\dagger_{\mu}( \mathbf q) \hat I^\dagger_{\nu}(- \mathbf q) \hat I^{\dagger N_{\mathrm{I}} - 1} \vert \Omega_{\mathrm{0}} \rangle$ as
\begin{equation}
    \Delta E_{\mathrm I} = - \xi \frac{e^2}{4 \pi \epsilon \epsilon_{\mathrm{0}} A} \mathcal I \alpha,
\end{equation}
with $A$ being the sample area, and parameter $\xi$ of the order of unity, $\xi \sim 1$, is tuned to match with experimental data setting an effective area. Here, $\alpha = (\alpha_{\mathrm{D}}^{-1} + \alpha_{\mathrm{I}}^{-1})^{-1}$ is the reduced particle Bohr radius. The dimensionless exchange integral $\mathcal I$ has the form
\begin{equation}
\begin{split}
    \mathcal I = & \int d^2x d^2y  \tilde V^{\mathrm{intra}}_{\mathrm{KR}}(\vert \mathbf x - \mathbf y \vert) \frac{1}{(1 + \alpha_1^2 x^2)^{3/2}} \frac{1}{(1 + \alpha_2^2 y^2)^3} \cdot
    \\
    &
    \cdot \left( \frac{1}{(1 + \alpha_1^2 x^2)^{3/2}} - \frac{1}{(1 + \alpha_2^2 y^2)^{3/2}} \right)
    \\
    & +
    \tilde V^{\mathrm{inter}}_{\mathrm{KR}}(\vert \mathbf x - \mathbf y \vert) \frac{1}{(1 + \alpha_2^2 y^2)^{3/2}} \frac{1}{(1 + \alpha_1^2 y^2)^{3/2}} \frac{1}{(1 + \alpha_2^2 x^2)^{3/2}} \cdot
    \\
    &
    \cdot \left( \frac{1}{(1 + \alpha_1^2 x^2)^{3/2}} - \frac{1}{(1 + \alpha_1^2 y^2)^{3/2}} \right),
\end{split}
\end{equation}
where $\tilde  V^{\mathrm{inter}}_{\mathrm{KR}}(\vert \mathbf x - \mathbf y \vert)$ is Keldysh-Rytova potential in dimensionless form, $\alpha_{\mathrm 1}$ and $\alpha_{\mathrm 2}$ are respectfully equal to $ \alpha_{\mathrm D}/ \alpha$ and $\alpha_{\mathrm I} / \alpha$. By estimating the exciton density as proportional to oscillator strength, we find $n_{\mathrm{I}} \simeq n_{\mathrm{D}} / 4$. Numerically, we estimate the shift with $\epsilon = 4$, $\alpha \simeq 6.7$ \AA. The resulting shift can be estimated as $\Delta E_{\mathrm{I}}=-0.24\xi~\mu$eV~$\mu$m$^2 \sqrt{n_{\mathrm{D}} n_{\mathrm{I}}}$. 



\bibliography{Manuscript_v9}
\end{document}